%% file: manuscript.tex
\documentclass[english,prb,aps,twocolumn,10pt]{revtex4-2}

\usepackage{amsmath}
\usepackage{amssymb}
\usepackage{bm}
\usepackage{graphicx}
\usepackage{babel}
\usepackage{array}
\usepackage{verbatim}
\usepackage[colorlinks=true, pdfstartview=FitV, linkcolor=blue, citecolor=blue, urlcolor=blue]{hyperref} 

\newcommand{\captionstyle}{\normalfont} 

\def\H        {{$^1$H \/}}
\def\Hns      {{$^1$H}}

\def\C        {{$^{13}$C \/}}
\def\Cns      {{$^{13}$C}}

\newcommand{\mr}[1]{\mathrm{#1}}
\newcommand{\unit}[1]{\,\mathrm{#1}}

\newcommand{\us}{\,\mu{\rm s}}

\newcommand{\ye}{\gamma_\mr{e}}
\newcommand{\yn}{\gamma_\mr{n}}

\newcommand{\cbracks}[1]{\ensuremath{\left\{#1\right\}}}
\newcommand{\pths}[1]{\ensuremath{\left(#1\right)}}
\newcommand{\bvec}[1]{\ensuremath{\mathbf{#1}}}

\newcommand{\sinf}[1]{\ensuremath{\sin\left(#1\right)}}

\newcommand{\NVa}{NV1 \/}
\newcommand{\NVans}{NV1}
\newcommand{\NVb}{NV2 \/}
\newcommand{\NVbns}{NV2}
\newcommand{\ImFFT}{\textrm{Im}[\textrm{FFT}]}
\newcommand{\ReFFT}{\textrm{Re}[\textrm{FFT}]}

\newcommand{\apar}{a_{||}}
\newcommand{\apari}{a_{||,i}}
\newcommand{\aperp}{a_{\perp}}
\newcommand{\aperpi}{a_{\perp,i}}

\newcommand{\Bo}{B_0}
\newcommand{\veca}{\vec a}
\newcommand{\vecai}{\vec a_i}

\newcommand{\vecBo}{\vec B_0}
\newcommand{\vecBcoil}{\vec B_\mr{coil}}

\newcommand{\vecex}{\vec e_x}

\newcommand{\vecez}{\vec e_z}
\newcommand{\ms}{m_S}
\newcommand{\po}{p_\mr{0}}
\newcommand{\poi}{p_{\mr{0},i}}
\newcommand{\rslice}{r_\mr{slice}}
\newcommand{\vecr}{\vec r}
\newcommand{\vecri}{\vec r_i}
\newcommand{\tbeta}{t_\beta}
\newcommand{\Te}{T_{2,\mr{e}}}
\newcommand{\Tn}{T_{2,\mr{n}}^\ast}
\newcommand{\tpol}{t_\mr{pol}}
\newcommand{\ts}{t_\mr{s}}
\newcommand{\tread}{t_\ell}

\newcommand{\betai}{\beta_i}
\newcommand{\betao}{\beta^\mr{(opt)}}
\newcommand{\phii}{\phi_i}

\newcommand{\hattheta}{\bm{\theta}}
\newcommand{\wi}{\omega_i}
\newcommand{\wo}{\omega_0}

\begin{document}

\title{Parallel detection and spatial mapping of large nuclear spin clusters}

\author{K. S. Cujia$^{1,2}$, K. Herb$^1$, J. Zopes$^{1,3}$, J. M. Abendroth$^1$, and C. L. Degen$^1$}
\affiliation{$^1$Department of Physics, ETH Zurich, Otto Stern Weg 1, 8093 Zurich, Switzerland.}
\affiliation{$^2$Present address: IT'IS foundation, Zeughausstrasse 43, 8004 Zurich, Switzerland.}
\affiliation{$^3$Present address: Institute for Biomedical Engineering, ETH Zurich, Gloriastrasse 35, 8092 Zurich, Switzerland.}
\email{degenc@ethz.ch}

\begin{abstract}
Nuclear magnetic resonance imaging (MRI) at the atomic scale offers exciting prospects for determining the structure and function of individual molecules and proteins.  Quantum defects in diamond have recently emerged as a promising platform towards reaching this goal, and allowed for the detection and localization of single nuclear spins under ambient conditions.
We present an efficient strategy for extending imaging to large nuclear spin clusters, fulfilling an important requirement towards a single-molecule MRI technique.  Our method combines the concepts of weak quantum measurements, phase encoding and simulated annealing to detect three-dimensional positions from many nuclei in parallel.
Detection is spatially selective, allowing us to probe nuclei at a chosen target radius while avoiding interference from strongly-coupled proximal nuclei.
We demonstrate our strategy by imaging clusters containing more than 20 carbon-13 nuclear spins within a radius of 2.4\,nm from single, near-surface nitrogen--vacancy centers at room temperature. The radius extrapolates to 7-8\,nm for \Hns.
Beside taking an important step in nanoscale MRI, our experiment also provides an efficient tool for the characterization of large nuclear spin registers in the context of quantum simulators and quantum network nodes.
\end{abstract}

\date{\today}

\maketitle


\section{Introduction}

Nuclear magnetic resonance (NMR) spectroscopy and magnetic resonance imaging (MRI) are powerful tools for molecular analysis and  and medical diagnostics.  While conventional NMR operates on millimeter-sized samples containing large ensembles of molecules, much effort has been directed at improving the resolution to the nanometer scale \cite{degen09,rose18} where the atomic structure could be analyzed at the level of individual molecules \cite{sidles91}.  Such a ``single-molecule MRI'' technique has the prospect of enabling direct imaging of molecular structures with three-dimensional resolution and chemical specificity.  This capability would lead to important applications in molecular biology, analytical chemistry, and many areas of nanoscale science and technology.

Quantum sensors based on nitrogen--vacancy (NV) centers in diamond have recently generated exciting progress in micron-scale \cite{glenn18,smits19} and nanoscale \cite{mamin13,staudacher13,loretz14apl} NMR spectroscopy.  Early experiments have demonstrated detection of single nuclear spins within the diamond crystal \cite{jelezko04nuclear,dutt07,dreau13} as well as of nanoscale films deposited on diamond surfaces \cite{mamin13,staudacher13,loretz14apl}.  Recent refinement of protocols has led to tremendous advances in sensitivity and spectral resolution \cite{boss17,schmitt17,glenn18}, allowing for the three-dimensional localization of individual nuclear spins \cite{zhao12,zopes18ncomm,zopes18prl,sasaki18}, spin pairs \cite{shi14,abobeih18,yang20}, and the chemical fingerprinting of molecular ensembles with high spectral resolution \cite{aslam17,glenn18}.  Most recently, the complete mapping of a 27-nuclear-spin cluster at cryogenic temperatures has been reported \cite{abobeih19}.

To extend experiments to the imaging of single molecules, methods are required that can efficiently detect and precisely localize a large number of distant nuclear spins in parallel.  To be compatible with single-molecule detection, the experimental arrangement requires very shallow defects ($\lesssim5\unit{nm}$) and preferably an ambient environment.
While advanced strategies have been proposed to solve the challenge of nuclear spin detection and localization \cite{ajoy15,kost15,perunicic16,wang16,schwartz19}, many of these strategies require very long coherence times or a single-shot readout of the quantum sensor to reach adequate sensitivity and spectral resolution \cite{abobeih19,bradley19}.  These conditions are difficult to realize with shallow defect centers at room temperature \cite{neumann10science,shields15}.  In addition, most proposed approaches require addressing nuclear spins one-by-one, leading to an unfavorable scaling as the number of nuclei become increasingly large.


In this work, we demonstrate a powerful method for the sensitive detection and spatial mapping of individual nuclei in large nuclear spins clusters.  Our approach combines the concepts of weak quantum measurements \cite{cujia19,pfender19}, phase encoding \cite{zopes18prl,sasaki18} and simulated annealing \cite{kirkpatrick83,tsallis96} to detect signal and extract precise three-dimensional distances from many nuclei in parallel.  We further show that our detection is spatially selective, allowing us to probe nuclei at a chosen target radius while avoiding interference from strongly-coupled proximal nuclei.
We demonstrate our strategy by mapping the \C environment of two NV centers containing 21 and 29 nuclei, respectively.  Because our experiments are performed on near-surface spin defects and at room temperature, they are compatible with the demanding environment of prospective single-molecule MRI investigations.
Besides taking an important step in developing a single-molecule MRI platform, our experiment also provides an efficient tool for the characterization of large qubit registers in the context of quantum simulators \cite{cai13}, quantum network nodes \cite{waldherr14,wolfowicz16,bradley19} and multi-qubit quantum processors \cite{krinner20,tsunoda20}.


\section{Imaging concept}

\begin{figure}[t]
\includegraphics[width=0.95\columnwidth]{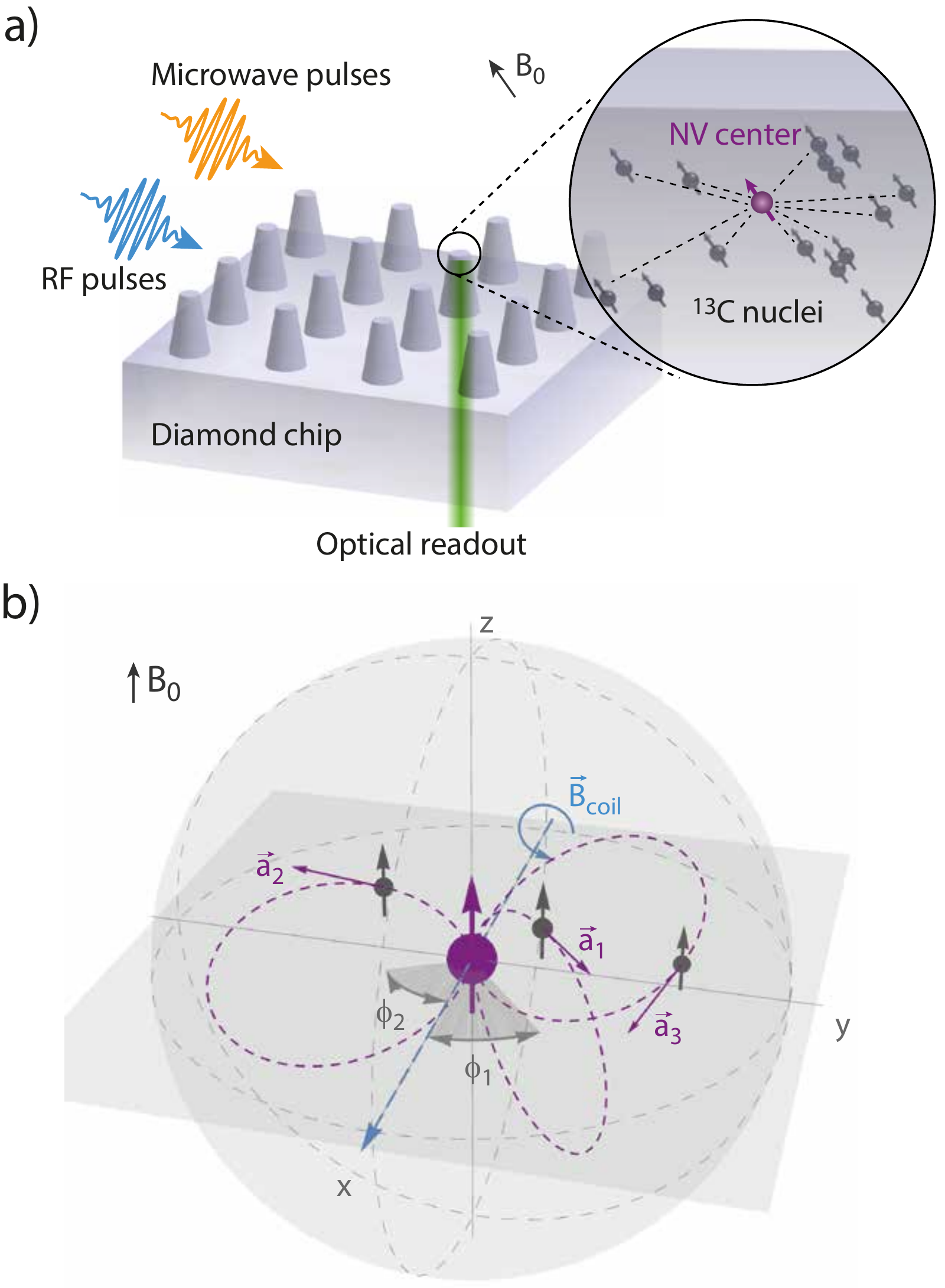}
\caption{\captionstyle
		(a) Concept of the nuclear spin mapping experiment.  We detect and image nuclear spins (black) surrounding a central electronic spin (purple).  In our experiments, spins are embedded in a nanostructured diamond chip, manipulated by microwave and radio-frequency (RF) pulses, and detected by optical means.
		(b) Spatial imaging is enabled through the hyperfine field $\veca = (\aperp\cos\phi,\aperp\sin\phi,\apar)$, whose magnitude and direction strongly depend on the three-dimensional position $\vecr=(r,\vartheta,\phi)$.  We determine the radius $r$ and polar angle $\vartheta$ by measuring the parallel and transverse hyperfine components, $\apar = \veca\cdot\vecez$ and $\aperp = |\veca\times\vecez|$ \cite{zhao12,boss16}.  The azimuth $\phi$ is equal to the phase of the nuclear precession \cite{zopes18prl,sasaki18}.
		$\vecez$ is a unit vector pointing along the electronic quantization axis and $\vecBo || \vecez$ is an external bias field.  $\vecBcoil || \vecex$ is the direction of the RF field.
    \label{fig1}
}
\end{figure}

Our concept and experimental situation is sketched in Fig.~\ref{fig1}a.  We consider a central electronic spin surrounded by a group of nuclear spins whose three-dimensional locations we aim to determine.  Here, both the electronic and nuclear spins are embedded in the solid matrix of a diamond crystal, but our concept is applicable to a general situation of a localized electronic spin \cite{christle15,iwasaki15,sukachev17} and a nearby nuclear ensemble, including surface molecules \cite{mamin13,staudacher13,lovchinsky16} or crystalline layers \cite{lovchinsky17}.
The electronic spin plays a dual role in our arrangement \cite{perunicic16}:  first, it acts as a local sensor for the weak magnetic fields produced by the nearby nuclei.  Second, it generates a strong magnetic dipole field that we exploit for spatial imaging.  In a reference frame where $z$ is the common quantization axis (Fig.  \ref{fig1}b), the dipole field is given by:
\begin{align}
\veca/\yn = \frac{\mu_0\hbar\ye\ms}{4\pi r^3} \left[ \frac{3\vecr(\vecez\cdot\vecr)}{r^2} - \vecez \right]  \ ,
\label{eq:veca}
\end{align}
where $\veca$ is the hyperfine vector (see Fig.~\ref{fig1}b), $\vecr = (r,\vartheta,\phi)$ are the polar coordinates of the nuclear spin relative to the electron spin situated at the origin, $\vecez$ is a unit vector along $z$, $\ms$ is the magnetic quantum number of the electronic spin ($\ms\in \cbracks{-1,0,1}$ for the NV center), and where we neglect Fermi contact effects \cite{zopes18ncomm}.  Further, $\mu_0$ is the vacuum permeability, $\hbar$ the reduced Planck constant, and $\ye$ and $\yn$ are the electronic and nuclear gyromagnetic ratios, respectively.  Thus, by measuring the three components of the hyperfine vector $\veca$, the distance vector $\vecr$ can be directly inferred (up to an inversion symmetry at the origin), revealing a spin's three-dimensional spatial location.

\section{Parallel signal acquisition}

\begin{figure}[t]
\includegraphics[width=0.95\columnwidth]{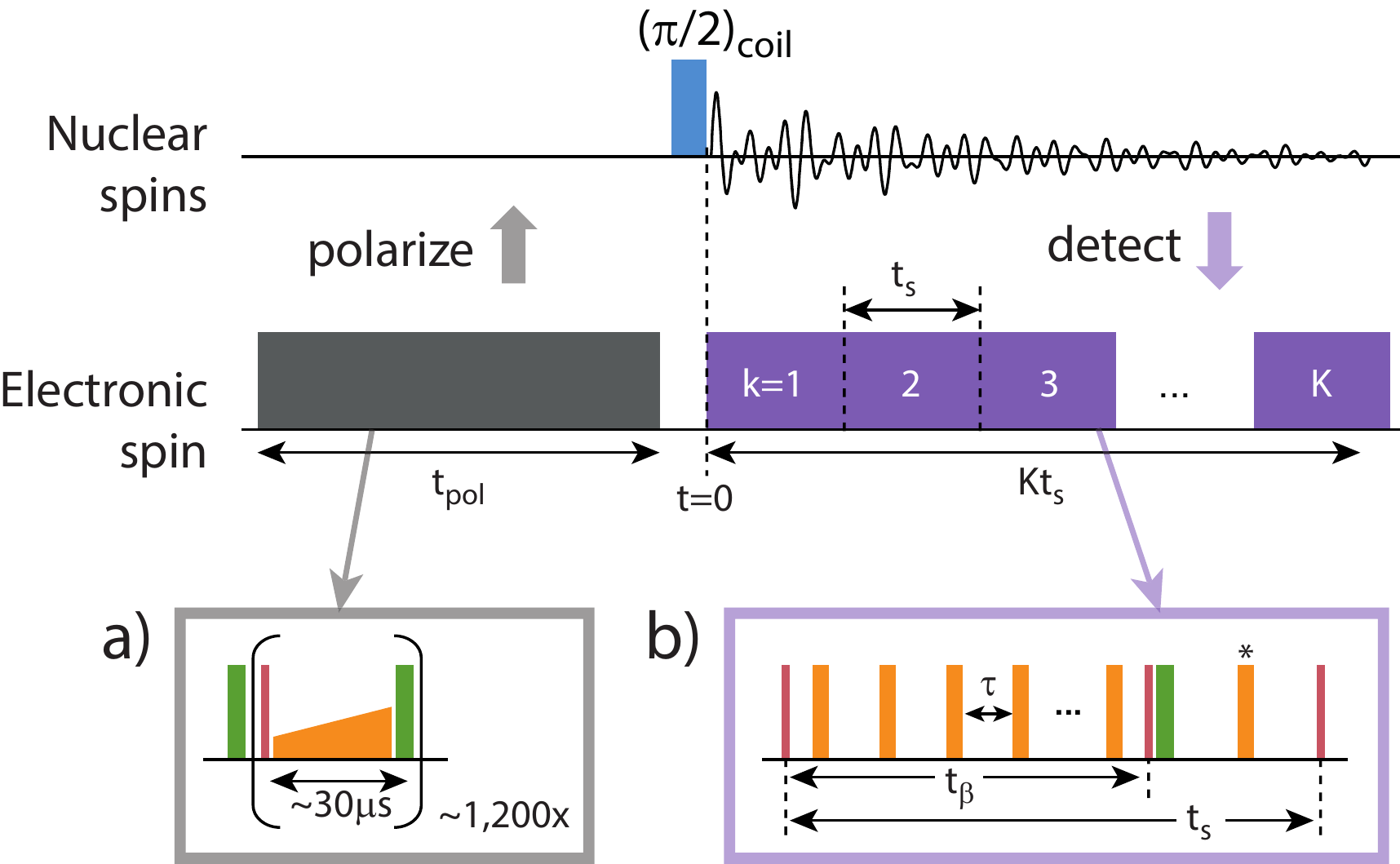}
\caption{\captionstyle
		Weak-measurement protocol for detecting the free precession signal from many nuclei in parallel.  We polarize the nuclear spins by a repeated NOVEL sequence (gray, inset a) \cite{henstra88,london13}, initiate simultaneous precession of all nuclei by applying a $\pi/2$ pulse with an external RF coil (blue) \cite{herb20}, and detect the precession by repeated sampling of the transverse nuclear magnetization (purple) \cite{boss17,schmitt17,degen17}. $\tpol$ is the polarization time, $\ts$ is the sampling time and $K$ is the number of samples.  Each weak-measurement read-out block (inset b) consists of a Carr-Purcell-Meiboom-Gill (CPMG) pulse train \cite{delange11,alvarez11,kotler11} of $4-24$ equidistant $\pi$ pulses (orange) separated by a delay time $\tau = 1/(2\yn\Bo)$, followed by an optical read-out pulse (green).  Red blocks are $\pi/2$ pulses.  The duration of the CPMG sequence defines the interaction time $\tbeta$.  An additional $\pi$ pulse ($\ast$) is used to average over the electronic $\ms=0,-1$ states.  See the Supplemental Material for experimental parameters \cite{supplemental}.
    \label{fig2}
}
\end{figure}

While three-dimensional localization has been demonstrated on individual nuclear spins \cite{zhao12,zopes18ncomm,zopes18prl,sasaki18}, the principal challenge lies in extending these experiments to large numbers of nuclei.  We address this challenge by exploiting the principle of weak quantum measurements \cite{colangelo17,cujia19}, which closely resembles the detection of a free-induction decay (FID) in canonical Fourier NMR spectroscopy.
Fig.~\ref{fig2} introduces our experimental protocol, consisting of a polarization, excitation and read-out step.  We begin by hyperpolarizing nuclear spins through a polarization transfer from the optically-aligned electronic spin (Fig.~\ref{fig2}a).  This initial step, when applied repetitively and for a sufficiently long time, leads to a volume of near-fully polarized nuclei around the central electronic spin \cite{broadway18}.
We then excite all nuclei simultaneously using a broad-band $\pi/2$ pulse and detect the free nuclear precession signal by sampling the transverse nuclear magnetization using weak measurements \cite{cujia19}.  The procedure yields an FID signal of the form:
\begin{align}
x(t) = \sum_{i=1}^{n} A(\beta_i) e^{-\Gamma(\beta_i) t}\cos[\wi t + \phi_i] \ ,
\label{eq:fid}
\end{align}
where $n$ is the number of nuclear spins.  Further, $A(\beta_i)$ is the probability amplitude \cite{boss17}, $\Gamma(\beta_i)$ the dephasing rate, $\wi$ the precession frequency, and $\phi_i$ the initial phase of the signal belonging to the $i$'th nucleus.
The parameter:
\begin{align}
\betai = \frac{\aperpi\tbeta}{\pi}
\label{eq:beta}
\end{align}
is a ``measurement gain'' parameter that is proportional to the hyperfine coupling constant $\aperpi$ multiplied by the interaction time $\tbeta$ of the ac detection (Fig.~\ref{fig2}b).  The parameter $\betai$, discussed in Section V,  plays an important role in single-spin FID detection as it governs the balance between signal gain and quantum back-action \cite{cujia19,pfender19,gefen18}.
We sample the FID at instances $t=k\ts$, where $\ts$ is the sampling time, $k=1...K$, and where $K$ is the number of points in the FID trace (Fig.~\ref{fig2}).

\section{Hyperfine parameters}

We next show that an FID trace described by Eq. (\ref{eq:fid}) contains all the information needed to reconstruct the hyperfine vectors $\vecai$, and hence, the three-dimensional locations $\vecri$ of the nuclear spins.
First, the parallel components $\apari$ (see Fig.~\ref{fig1}b) can be determined from the spectral positions of the nuclear resonances, given by the free precession frequencies:
\begin{align}
\wi
	&= \frac12\left( \wo + \sqrt{(\wo+\apari)^2 + \aperpi^2} \right)
	\approx \wo + \frac12 \apari  \ ,
	\label{eq:apar}
\end{align}
where the approximation is for small $\apari,\aperpi\ll \wo$ (fulfilled in our experiments) and where $\wo = -\yn B_0$, with $B_0||\vecez$ being the external magnetic bias field~\cite{boss16}.  Because the $\apari$ values of spins are in general different, they provide a means to separate out nuclear signals in the Fourier spectrum. (Accidental overlap of resonances could be resolved using two-dimensional NMR schemes \cite{boss16}).

Second, the amplitudes $A(\betai)$ and decay rates $\Gamma(\betai)$ encode information about the perpendicular components $\aperpi$:
\begin{subequations} 
\label{eq:aperp}
\begin{align}
A(\beta_i)      &= \frac12 \poi \sinf{\betai} \approx \frac{\po\aperpi\tbeta}{2\pi} \label{eq:a} \ , \\
\Gamma(\beta_i) &= \frac{\aperpi^2\tbeta^2}{4\ts\pi^2} + \frac{1}{\Tn} + \frac{\apari^2\tread^2}{2\ts} \label{eq:gamma} \ ,
\end{align}
\end{subequations}
where $\poi$ is the initial polarization of the $i$'th nuclear spin.
We will assume that all nuclei carry approximately the same polarization ($\poi\approx\po$); a justification is given in Refs. \onlinecite{broadway18,supplemental}.
Note that $p_0$ also contains any pulse errors and other imperfections of our pulse sequence, and therefore rather reflects a pre-scaling factor and lower bound for the nuclear polarization. 
The dephasing rate of nuclei is influenced by three effects: (i) a measurement-induced dephasing proportional to $\aperpi^2\tbeta^2$ due to quantum backaction \cite{cujia19}, (ii) an intrinsic $T_2^\ast$ dephasing (assumed the same for all nuclei \cite{supplemental}), and (iii) an additional decay rate proportional to $\apari^2\tread^2$ that is specific to the stochastic optical read-out process of the NV center with effective duration $\tread$ \cite{supplemental,cujia19}.

Finally, the azimuth $\phii$ is encoded in the complex phase of the nuclear FID signal.  Because we initiate the FID by applying a $\pi/2$ pulse with an external RF coil, all nuclei are rotated around a common laboratory-frame axis and start precession with the same phase.  By contrast, the ac detection of the FID is phase-sensitive with respect to each nucleus' individual hyperfine field.  As a consequence, the phase $\phi_i$ is equal to the spatial angle between the coil and hyperfine axes in the laboratory frame (Fig.~\ref{fig1}b) \cite{zopes18prl,sasaki18}.  Analysis of the complex FID signal therefore directly reveals the desired azimuth $\phi_i$.

\section{Sensitive slice}
\label{sec:sensitiveslice}
\begin{figure}[t]
\includegraphics[width=0.99\columnwidth]{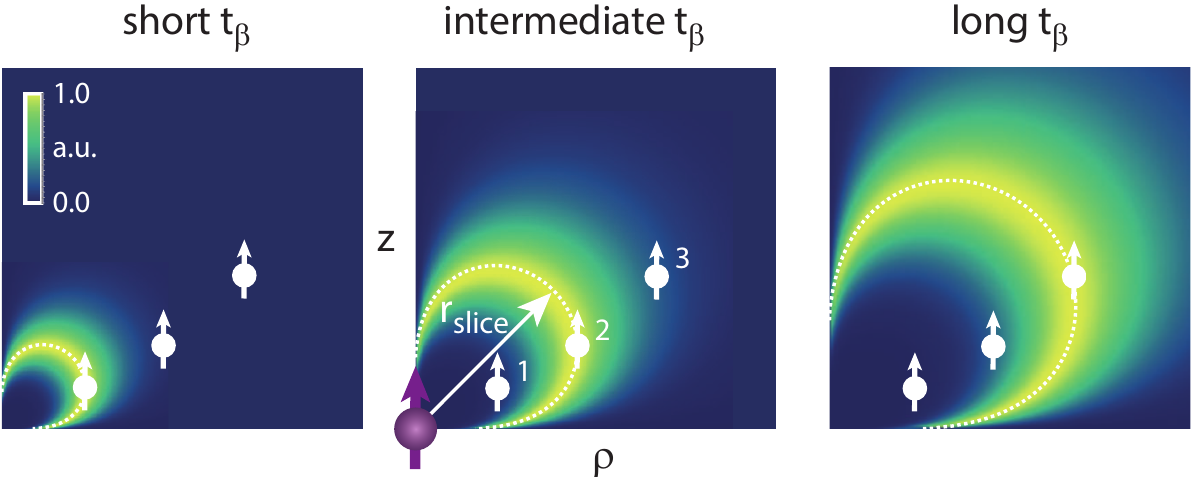}
\caption{\captionstyle
		Simulation of the sensitive slice $\mathcal{S}(\vecr)$ for different interaction times $\tbeta$ shown in a $\rho z$ plot, where $z = r\cos\vartheta$ and $\rho = r\sin\vartheta$ are vertical and radial distance, respectively.  Bright shading color codes the signal-to-noise ratio.  By varying $\tbeta$ the sensitive slice can be tuned to spins in close (1), intermediate (2) or far (3) distance from the central electronic spin (panels from left to right).  The dashed contour is the $\aperp$ isoline where $\beta=\betao$.  $\rslice$ is the characteristic radius of a slice.	 Plots assume $\tread=0$ and $(\Tn)^{-1}=0$.
    \label{fig3}
}
\end{figure}
The magnitude of the FID signal strongly depends on a spin's three-dimensional position $\vecr$, because of the position dependence of the hyperfine interaction.  We can capture the spatial dependence by calculating a sensitivity function $\mathcal{S}(\vecr)$ that quantifies the signal contribution as a function of spin location $\vecr$.  The sensitivity function is expressed as a signal-to-noise ratio \cite{supplemental} and given by:
\begin{align}
\mathcal{S}(\vecr) \equiv \mathcal{S}(\beta(\vecr))
= \frac{A(\beta)}{\Gamma(\beta)\ts\sqrt{K}}\frac{\pths{1-e^{-\Gamma(\beta) K\ts}}}{\sqrt{\tpol+K\ts}}
\label{eq:snr}
\end{align}
where $\beta = \aperp\tbeta/\pi$ (Eq. \ref{eq:beta}) encodes the spatial position (via the hyperfine parameter $\aperp$),  and where $K$, $\ts$, $\tbeta$ and $\tpol$ are experimental parameters defined in Fig. \ref{fig2}.

Fig. \ref{fig3} plots $\mathcal{S}(\vecr)$ as a function of vertical and radial distance to the central electronic spin.  Interestingly, the sensitivity does not monotonically decay with distance, as might be expected from the $\aperp \propto r^{-3}$ scaling of the hyperfine interaction.  Rather, $\mathcal{S}$ is initially low, and increases with $r$ until it reaches a maximum at a characteristic radius $\rslice$ before showing the expected $r^{-3}$ decay.  The suppression of signal from close spins is a consequence of quantum back-action \cite{cujia19}:  Because these spins are strongly coupled, their measurement strength parameter $\beta$ is large, leading to a rapid signal decay $e^{-\Gamma(\beta) t} \rightarrow 0$ (Eq. \ref{eq:gamma}).  Conversely, distant spins with small $\beta$ generate weak signals because $A(\beta)\rightarrow 0$ (Eq. \ref{eq:a}).  Maximum sensitivity results at an intermediate value where the two effects are balanced,
\begin{align}
\betao \approx \sqrt\frac{2}{K} \ .
\label{eq:betao}
\end{align}
The optimum point of sensitivity is reached when intrinsic and induced decay rates are commensurate, $(\Tn)^{-1} = \beta^2/(4\ts)$ and when the FID record length is matched to the decay rate, $K\ts = 1/\Gamma$ \cite{supplemental}.

As shown in Fig. \ref{fig3}, the points of maximum sensitivity are located along a contour of constant $\aperp = \pi\betao/\tbeta$.  We denote this contour the ``sensitive slice'' associated with the interaction time $\tbeta$.  By varying $\tbeta$, we can vary the radius of the sensitive slice $\rslice$ and tune detection from close to distant nuclear spins (Fig. \ref{fig3}, left to right).  Because $\aperp \propto r^{-3}$, the radius of the sensitive slice scales as $\rslice\propto \tbeta^{1/6}$ \cite{supplemental}.
The spatial selectivity in an important feature of our method, since it allows us to selectively probe nuclear spins at a defined (far) distance from the central electronic spin while avoiding interference from strongly-coupled nuclei in close proximity.  Further, by sweeping $\tbeta$, we can collect FID traces from several sensitive slices and cover an extended spatial volume in the sample.

\section{Maximum Likelihood Estimation by Simulated Annealing}
\begin{figure*}[t]
\includegraphics[width=0.8\textwidth]{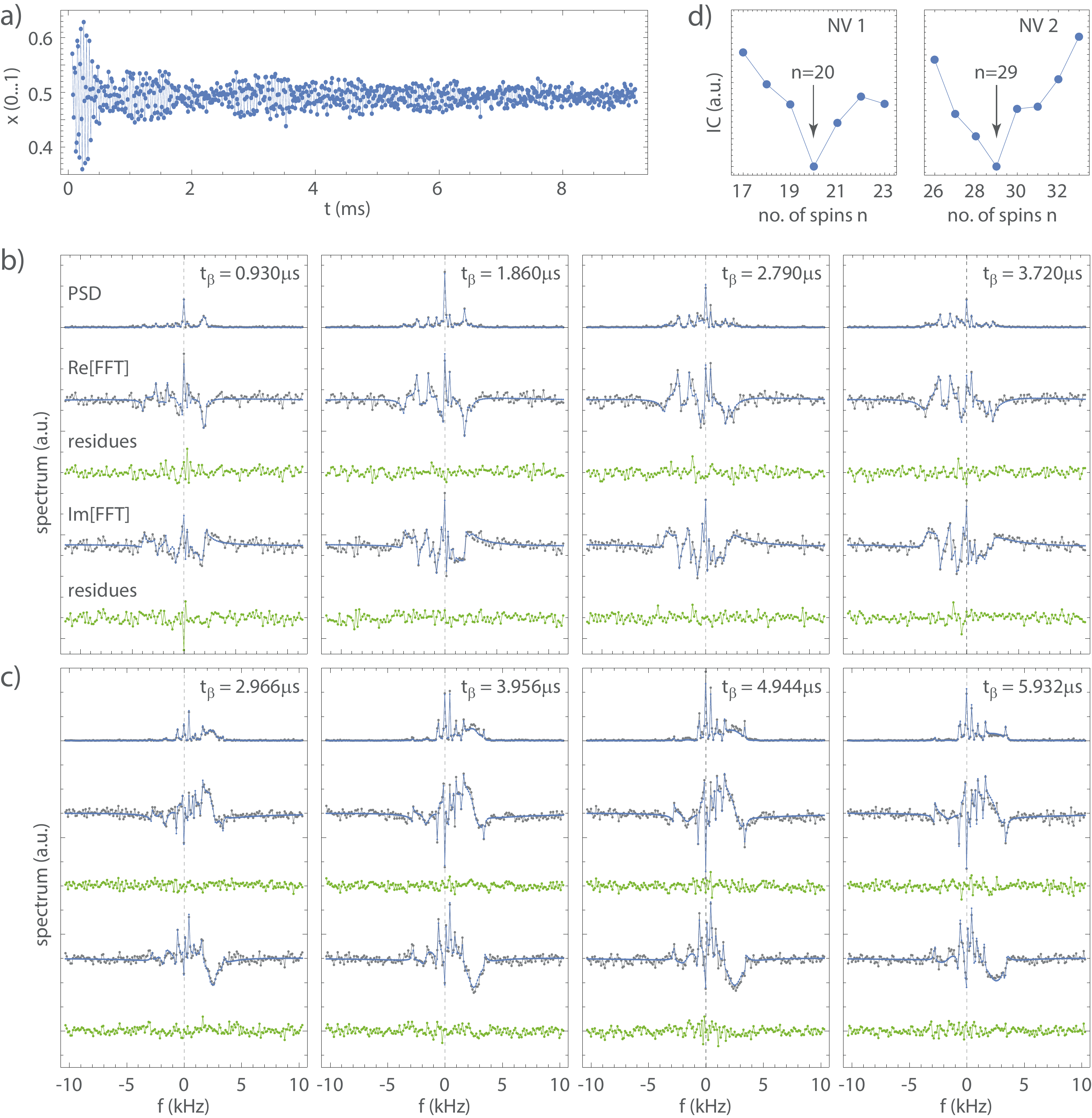}
\caption{\captionstyle
    Experimental demonstration of \C NMR spectroscopy of large spin clusters. 
    (a) Example FID trace from \NVbns.  Each data point reflects the probability amplitude $x(t)$ from one read-out block, integrated over $\sim 10^6$ repetitions of the full sequence (see Fig. \ref{fig2}).  Measurement parameters are $\tpol=40\unit{ms}$, $\ts=11.48\unit{\us}$, $\tbeta=4.944\unit{\us}$ and $K=800$, and total measurement time is $11\unit{h}$. 
    (b) Complex Fourier spectra of the \C environment of \NVa for a series of interaction times $\tbeta$.  Shown are from top to bottom (vertically offset for clarity): power spectrum (PSD), real part of the complex spectrum ($\ReFFT$), fit residues for $\ReFFT$, imaginary part of the complex spectrum ($\ImFFT$), and fit residues for $\ImFFT$.  Blue traces are best fits (see text).  The horizontal axis shows the spectral shift relative to the \C Larmor frequency at $2.156\unit{MHz}$ calibrated using correlation spectroscopy \cite{boss16}.  The bias field is $B_0 = 201.29\unit{mT}$.
    The PSD data of \NVa are the same as in Ref. \onlinecite{cujia19}, Extended Data Fig.~5a.
	(c) Fourier spectra of the \C environment of \NVbns.  Bias field is $B_0=188.89\unit{mT}$.
	(d) Cost function [Eq. (\ref{eq:ic})] plotted as a function of the number of spins $n$.  Best fits are obtained for $n=20$ (\NVans) and $n=29$ (\NVbns).
    \label{fig4}
}
\end{figure*}
Armed with a protocol for measuring the signals and coupling constants from many nuclei in parallel, we develop a maximum likelihood protocol to extract the hyperfine parameters and position vectors from an FID trace (Eq. \ref{eq:fid}).  Assuming $n$ spins are contributing to the signal, our model contains $M=3n+3$ unknown parameters, including the three hyperfine parameters $\apari$, $\aperpi$ and $\phii$ for each spin~$i$ plus three additional, global parameters accounting for an initial polarization $p_0$ and dephasing times $\Tn$ and $\tread$  [see Eq. (\ref{eq:aperp})].  To proceed, we collect the unknown parameters in a single parameter vector $\hattheta = \cbracks{\theta_m}$, where ${m=1\dots M}$.  Our goal is to balance goodness of the fit and model complexity by minimizing a cost function of the form:
\begin{align}
\mr{IC} = G(\hattheta,\bvec{x}) + P(K,M) \ ,
\label{eq:ic}
\end{align}
where $G(\hattheta,\bvec{x})$ is a measure of the goodness of the fit, $\bvec{x} = \cbracks{x_k}$, where ${k=1\dots K}$, is the set of measured data points, and $P(K,M)$ is a penalty term to prevent over-fitting \cite{supplemental,knowles16}.
Eq. \ref{eq:ic} is the generic form of a so-called information criterion (IC).
In a likelihood framework, $G(\hattheta,\bvec{x})$ can be expressed in terms of a negative likelihood function \cite{burnham02}:
\begin{align}
G(\hattheta,\bvec{x})
  = - 2\ln\left( \mathcal{L}[\hattheta,\bvec{x}]\right)
	= K\ln\pths{\sum_{k=1}^{K}[x_k - \tilde{x}_k(\hattheta) ]^2 },
\label{eq:wic_neglog}
\end{align}
where the argument of the logarithm is the residual sum of squares.  The function $\tilde{x}_k(\hattheta)$ represents the estimated data points calculated from Eq.~(\ref{eq:fid}) using the parameter vector $\hattheta$.
The most common penalty terms $P(K,M)$ are defined by the Akaike Information Criterion (AIC) \cite{akaike74} and the Bayesian Information Criterion (BIC) \cite{schwarz78}.  In our strategy, we combine the AIC and BIC penalties into a single term, known as the Weighted Average Information Criterion (WIC) \cite{wu98} and defined in the Supplemental Material \cite{supplemental}, that combines the strengths of both criteria and performs well independent of sample size $K$.

By minimizing Eq.~(\ref{eq:ic}) we obtain the most likely parameter vector $\hattheta$ and the vector size $M$, which reflects the minimum number of spins $n$ needed to explain the data.  The minimization is challenging due to the large number of unknown parameters $M$.
We implement the method of generalized simulated annealing (GSA) to address this challenge.  GSA is a stochastic approach \cite{tsallis96} that combines the original method of classical simulated annealing \cite{kirkpatrick83} and fast simulated annealing \cite{szu87}.  The method has proven especially useful for global optimization of complicated, multi-dimensional systems with large numbers of local minima such as those present in quantum chemistry, genetics or non-linear time series \cite{xiang97}.  Due to its statistical nature, local minima can be escaped much more easily than with steepest-descent or gradient methods. The core idea is to combine importance sampling with an artificial temperature which is gradually decreased to simulate thermal noise.
To improve the GSA, we run the minimization over a large number of random starting configurations for $\hattheta$.  Finally, once a best-estimate set of parameters has been found, we can compute the three-dimensional locations $\vecri$ of nuclei from the hyperfine vector $\vecai$ by inverting Eq.~(\ref{eq:veca}) (see Appendix A).


\section{Experimental demonstration}
\begin{figure*}[t]
\includegraphics[width=0.99\textwidth]{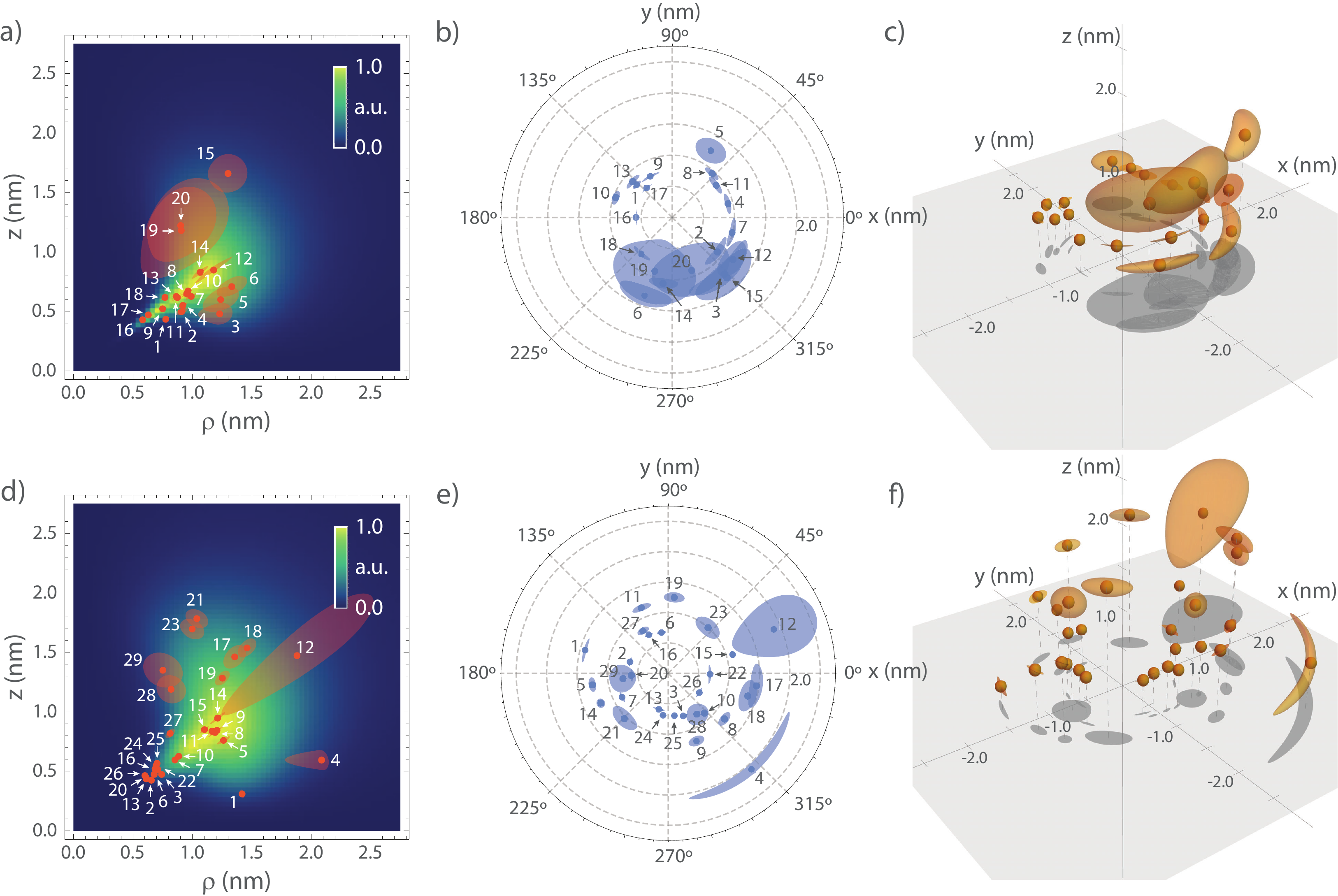}
\caption{\captionstyle
    Visualizations of the three-dimensional nuclear spin mapping.
		(a) Fitted spatial locations of the \C spins of \NVa (dots) shown in a $\rho z$-plot.  The central NV spin is located at the origin.  Note that our method is ambiguous with respect to an inversion at the origin.  Therefore, all nuclei are plotted in the upper hemisphere.  Color coding reflects the combined sensitive slice [Eq. (\ref{eq:snr})] taking all $\tbeta$ values as well as finite $\tread$ and $\Tn$ into account.
		(b) Polar plot showing the azimuth $\phi$ on the $xy$-plane.
		(c) Three-dimensional view of the nuclear positions.
		(d-f) Corresponding plots for the carbon spins of \NVbns.
		Uncertainties ($\pm$1 standard error) are shown as shaded areas or volumes, respectively, and are omitted if smaller than the data points.  Index labels refer to Tables S3 and S4 in the Supplemental Material \cite{supplemental}.
    \label{fig5}
}
\end{figure*}
We experimentally demonstrate our three-dimensional nuclear localization strategy by imaging the \C environment of shallow NV centers in diamond.  We focus on two NV centers in this work, labeled \NVa and \NVbns, out of $5$ recorded datasets.   The two NV centers are selected for favorable optical contrast and electron spin coherence times, but not for their \C environment.  Their shallow depth ($\sim 10\unit{nm}$) is not important for this study except for demonstrating that our method is compatible with near-surface NV centers.  We probe the NV centers at room temperature using non-resonant optical excitation and a single-photon counting module \cite{supplemental}.  Electronic and nuclear spins are manipulated via two arbitrary waveform generators connected to separate microwave transmission line and RF micro-coil circuits, respectively \cite{zopes18prl,herb20}.  Experiments use a bias field $B_0 \sim 200\unit{mT}$ aligned to within $1^\circ$ of the NV symmetry axis (Fig.~\ref{fig1}b).


Fig.~\ref{fig4}a shows an example of an FID time trace from \NVb for $\tbeta=4.944\unit{\us}$, and Fig.~\ref{fig4}b,c show the complete data set of \C Fourier spectra obtained for both NV centers.  For each NV center, we record four spectra with different values of the interaction time $\tbeta$ to sample different radii of the sensitive slice and to add redundancy.  For each dataset, we plot the power spectrum, the real and imaginary parts of the complex Fourier spectrum, as well as the fit residues.  Clearly, the spectra show a rich peak structure, indicating that we are detecting a large number of \C resonances. 

To extract the hyperfine parameters, we apply Eq. (\ref{eq:ic}) to the combined set of complex Fourier spectra.  We fit the spectra, rather than the FID traces, to improve the robustness of the fit (see below).  We begin by randomly initializing each parameter, and then minimize the residues between the experimental and computed spectra using a GSA algorithm \cite{supplemental} on a high-performance computer cluster \cite{euler}.  To improve the robustness of the fit, we compute separate residues for real and imaginary parts of the complex spectrum as well as for their magnitude squared, and minimize the sum of all residues within the likelihood framework \cite{supplemental}.  Further, we penalize configurations where the distance between any two \C is less than one bond length.
To accelerate the search for a global minimum, we repeat the procedure for a large number ($\sim 10^2$) of starting values randomly chosen from pre-defined parameter intervals \cite{supplemental}.
Finally, to determine the number of spins we run the minimization routine for different $n$ and select the configuration with the smallest global IC value (Fig.~\ref{fig4}d).
Once the minimization has terminated and $n$ has been determined, we perform bootstrapping on the final fit residues \cite{efron79,supplemental} to obtain an estimate for the fit uncertainties for all parameters. 

The calculated spectra for the most likely nuclear configurations are displayed as blue solid lines in Fig.~\ref{fig4}b,c.  For the two datasets we find $n=20$ for \NVa and $n=29$ for \NVb (Fig.~\ref{fig4}d).  The good agreement with the experimental data (gray dots) and the small residues (green), which are of the same order as the measurement noise, indicate that our model and fitting method are appropriate.  We have verified the calculated spectra by performing a full density matrix simulation using the final parameter set (Fig. S2 \cite{supplemental}).  All fit results are collected in Tables S3 and S4 \cite{supplemental}.

Fig.~\ref{fig5} shows visualizations of the three-dimensional locations of nuclei.  We find that nuclear positions are clustered between ca. $r$=0.7-2.4\,nm and ca. $\vartheta$=30-75$^\circ$.  This clustering is a consequence of the spatial selectivity of our method:  although the \C nuclei are distributed randomly over the diamond lattice, only spins falling within the sensitive slice are picked up by the weak measurement detection sequence.  The combined sensitive slice for all measurements, taking the $\tbeta$ values of the four spectra, $\tread$ and $\Tn$ dephasing into account, agrees well with the extracted \C positions.  (The $\tread$ dephasing suppresses signal from nuclei with large $\apar$, leading to a low sensitivity for spins near $\vartheta = 0^\circ$ and $90^\circ$).  Fig.~\ref{fig5} also clearly shows that the spatial precision of our method is highest and well below 1\,\AA\ for \Cns's that are located near the maximum of the sensitive slice, while the precision can be poor for spins located at the fringe of the slice.

Due to the large parameter space and complex structure of the minimization problem it is not guaranteed that the solution presented in Fig. \ref{fig5} represents the ``true'' configuration of \C nuclei.  In fact, our method may report a single-spin position that in fact represents a pair or cluster of \Cns.  This issue could be resolved my measuring nuclear spin-spin interactions using two-dimensional NMR (see below).
While we do not have independent means for verifying the three-dimensional nuclear configuration, we can perform some basic statistical tests on the density, spatial distribution and fit uncertainty of \C positions.
Comparing the volume uncertainties $\delta V$ of the \C (indicated by orange shading in Figs. \ref{fig5}c,e and tabulated in \cite{supplemental}) with the volume per carbon atom in the diamond lattice ($V=5.67\unit{\AA^3}$ \cite{shikata18}), we find that 13 out of 20 spins (\NVans) and 21 out of 29 spins (\NVbns) have $\delta V<V$ and therefore likely represent single nuclei. 
Next, defining the volume of the sensitive region by the volume in space contributing 50\% to the total signal~\cite{mamin13}, we find sensitive volumes of $V=9.3\unit{nm}^3$ and $14.8\unit{nm}^3$ for \NVa and \NVbns, respectively.  Considering an average density $\rho(^{13}\text{C}) = 1.94\unit{nm^{-3}}$ for \C nuclei in diamond at natural isotope abundance (1.1\%), the average number of \C in the sensitive slices are 18.0 and 28.6, respectively, in good agreement with our experimental result.
Further, a $\chi^2$ test for the angular distributions of the azimuth angles yields $p$-values well above the $5\%$ level ($54\%$ and $70\%$, respectively), as expected for a random \C distribution.  Within these statistics, our experimental results are fully consistent with a stochastic distribution of \C atoms around the NV centers.

\section{Outlook}

Before concluding, we point out some limitations of our present experiment and how these may be resolved in the future.
A central issue is that there is no direct way of confirming the maximum likelihood solution beyond testing the robustness of the minimization.  In addition, because of the inversion symmetry of the dipolar interaction [Eq. (\ref{eq:veca})], our method does not discriminate between spins lying in the upper and lower hemisphere.
The logical way forward is to introduce additional spatial constraints.  The inversion ambiguity can be resolved, for example, by positioning all spins in one hemisphere.  This is a natural arrangement for single-molecule detection or planar spin registers.  Another option is the application of a strong external magnetic gradient \cite{grinolds14}.

The most promising avenue, however, is the inclusion of nuclear spin-spin interactions.  Such pair-wise nuclear spin couplings can provide precise local constraints \cite{abobeih19} that ideally complement the global distance information provided by our method.  While internuclear couplings between \C are small in our experiment because of the low isotope abundance \cite{supplemental}, couplings between protons in molecules can exceed $10\unit{kHz}$ and dominate over the hyperfine interaction. 
The two-dimensional (2D) NMR techniques needed to perform internuclear distance measurements, like nuclear-Overhauser-effect or electron-spin-echo-envelope-modulation spectroscopies, can be conveniently integrated with our protocol because of its similarity to conventional FID detection.
While multi-dimensional spectroscopy schemes will further increase the measurement time and complexity for retrieving three-dimensional structures, techniques exists to address these challenges, such as sparse sampling \cite{mobli12,kost15} and machine learning \cite{aharon19,kong20,jung21}.

\section{Summary}

By demonstrating three-dimensional mapping of nuclear spins in ambient conditions, our work takes an important step forward towards the ambitious goal of single-molecule MRI.  Next steps include extending the sensing radius of the NV center to $\gtrsim 4\unit{nm}$, which is the minimum depth reported for stable near-surface NV centers \cite{sangtawesin19}.  While the most distant \C in this work is at $r\approx 2.4\unit{nm}$ (\C no. 12 in Fig.~\ref{fig5}d), increasing the interaction time to the coherence time ($\Te$=50-200$\unit{\us}$ \cite{supplemental,sangtawesin19}) will scale this distance to $r\sim 4.5\unit{nm}$ for \C and to $r \sim 7-8\unit{nm}$ for the high-moment nuclei \H and $^{19}$F, respectively.  Reaching this limit will require very high magnetic field stability, and work is currently underway to address this challenge.  Another step includes the controlled termination and subsequent functionalization of the diamond surface to bind desired target molecules and simultaneously decouple the shallow NV center from the surface chemistry \cite{kaviani14,kawai19}.

Beside magnetic imaging of single spins, our work also provides interesting perspectives for the characterization of large qubit registers in quantum applications.  For example, our method can be applied to efficiently map out the coupling network of quantum nodes built from a central electronic spin backed by a nuclear spin register \cite{awschalom18,zwanenburg13}.  Such quantum nodes are central elements in emerging optical \cite{bernien13} or electronic \cite{tosi17} quantum interconnects.  Another application is nuclear quantum simulation using an electronic qubit for initialization and readout \cite{cai13}.  Finally, our parallel measurement protocol could provide a rapid means for calibrating cross-talk in superconducting qubit architectures \cite{krinner20,tsunoda20}.


\section*{Acknowledgements}

We thank M. L. Palm for helpful discussions and acknowledge technical support from the machine shop of the Department of Physics at ETH Zurich.
This work has been supported by Swiss National Science Foundation (SNFS) Project Grant No. 200020\_175600, the National Center of Competence in Research in Quantum Science and Technology (NCCR QSIT), and the Advancing Science and TEchnology thRough dIamond Quantum Sensing (ASTERQIS) program, Grant No. 820394, of the European Commission.



\appendix

\section{Hyperfine vector $\veca$ and position $\vecr$}

The radius $r$ and polar angle $\vartheta$ are computed from the parallel and transverse hyperfine parameters $\apar$ and $\aperp$ as follows (see Ref. \onlinecite{boss16}, Eq. S28 and S29):
\begin{align}
\vartheta &= \arctan\left\{ \frac{1}{2}\left( -3\frac{\apar}{\aperp} + \sqrt{9\frac{\apar^2}{\aperp^2}+8} \right)\right\} , \label{eq:theta} \\
r &= \left\{ \frac{\mu_0 \ye\yn\hbar(3\cos^2\vartheta-1)}{4\pi\apar} \right\}^{1/3} \label{eq:r}.
\end{align}

\input{"references.bbl"}

\end{document}



\begin{center}

\large
\vspace{15 mm}
\textbf{{Supplemental Material for: \\ \vspace{10 mm} ``Parallel detection and spatial mapping\\ of large nuclear spin clusters''}}

\normalsize
\vspace{15 mm}
K. S. Cujia$^{1,2}$, K. Herb$^1$, J. Zopes$^{1,3}$, J. M. Abendroth$^1$ and C. L. Degen$^1$

\vspace{5 mm}
\textit{$^1$Department of Physics, ETH Zurich, Switzerland.}\\
\textit{$^2$Present address: IT'IS foundation, Zeughausstrasse 43, 8004 Zurich, Switzerland.}\\
\textit{$^3$Present address: Institute for Biomedical Engineering, ETH Zurich, Gloriastrasse 35, 8092 Zurich, Switzerland.}

\end{center}


\clearpage
\section{Materials and Methods}

\subsection{Diamond samples}
Two single-crystal diamond plates were used for experiments.
Both sample A (NV1) and sample B (NV2) were electronic-grade, natural abundance (1.1\% \Cns) diamond membranes.
NV centers were created by $^{15}$N$^+$ ion implantation at an energy of $5\unit{keV}$ and doses of $5\cdot 10^{11}\unit{cm}^{-2}$ and $4\cdot 10^{10}\unit{cm}^{-2}$ for samples A and B, respectively.  Samples were subsequently annealed at $850\unit{^{\circ}C}$ to form NV centers.  We chose the $^{15}$N species to discriminate implanted NV centers from native ($^{14}$N) NV centers. Both samples were cleaned in a 1:1:1 mixture of H$_2$SO$_4$:HNO$_3$:HClO$_4$ and baked at $465\unit{^{\circ}C}$ in air before mounting them in the setup.  Whenever organic contamination was spotted, samples were cleaned in a 2:1 mixture of H$_2$SO$_4$:H$_2$O$_2$ (Piranha).
We etched nano-pillars into the membrane surfaces to increase the photon collection efficiency. The continuous wave (CW) photon count rate was $250-500\unit{kC/s}$.

\subsection{Experimental setup}
Experiments were performed using a custom-built confocal microscope equipped with a green $\lambda=532\unit{nm}$ frequency-doubled Nd:YAG excitation laser (CNI Laser MSL-FN-532nm) and a $630-800\unit{nm}$ detection path using a single-photon avalanche photo diode (APD, Pelkin Elmer SPCM-AQR Series). Optical pulses were generated by an acousto-optic modulator (AOM, Crystal Technology 3200-144) in a double-pass configuration and gating of arriving photons was realized by time-tagging (NI-PCIe-6363) and software binning of photon counts. Typical laser excitation powers were on the order of $100\unit{\mu W}$.

We synthesized microwave pulses for manipulating the electronic spin using an arbitrary waveform generator (AWG, Tektronix AWG5012C) and up-converted them to $\sim 2.5\unit{GHz}$ using a local oscillator (Hittite HMC-T2100) and a quadrature mixer (Marki microwave IQ1545).  Pulses were subsequently amplified (Gigatronics GT-1000A) prior to delivery to the NV center using a coplanar waveguide (CPW) photo-lithographically defined on a quartz cover-slip.  The transmission line was terminated on an external 50$\,\Omega$ load (Meca 490-2).
We synthesized radio-frequency (RF) pulses for nuclear spin manipulation using an AWG (National Instruments PCI-5421) and subsequently amplified them (Mini-Circuits LZY-22+). The pulses were transmitted using a planar micro-coil connected in series with a $50\Omega$ termination (Meca 697-30-1). The measured micro-coil inductance was $L = 0.77 \unit{\mu H}$. The $50\Omega$ termination increased the rf-circuit bandwidth ($Q = L/R$) at the expense of efficiency (most power was dissipated in the load). The micro-coil circuit had a $3\unit{dB}$-bandwidth of $\sim 19\unit{MHz}$. \C Rabi frequencies were typically around $25\unit{kHz}$.  A layout of the micro-coil arrangement is given in Ref. \onlinecite{herb20}.

We used a cylindrical samarium-cobalt permanent magnet (TC-SmCo, reversible temperature coefficient $0.001\%/\degree\mr{C}$) to create a bias field $B_0\sim 190\unit{mT}$ at the NV center location. To align $B_0$ with the NV symmetry axis, we adjusted the relative location of the permanent magnet by fitting to a set of electron paramagnetic resonance (EPR) lines recorded at different magnet locations and subsequently by maximizing the CW photon count rate.

\subsection{Tracking of magnetic field drifts}
The net magnetic bias field drifted by typically a few Gauss, leading to variations in the EPR frequency of order $\pm 100\unit{kHz}$ and variations in the \C Larmor frequency of order $\pm 50\unit{Hz}$.
We continuously tracked and logged magnetic field drifts by measuring the EPR resonance of the NV center and periodically re-adjusted the microwave excitation frequency during the course of the experiment.
%
%
\section{Maximum likelihood estimation by simulated annealing}

\subsection{Information criteria}
%
The selection of appropriate approximate models is critical for statistical inference out of experimental data. A very general methodology for model selection and parameter estimation is the use of information criteria and likelihood concepts \cite{burnham02}.
When applying information criteria for model selection one aims to measure the distance or information, which in turn can be linked to the concept of entropy maximization, between two models \cite{burnham02}.

From a simplified perspective, the idea is to balance the goodness of fit and the complexity of a model using a cost function of the form
\begin{align}
    \mathrm{IC} = G\pths{\sigma^2} + P\pths{K, M} \ , 
    \label{eq:information_criteria}
\end{align}
where $G\pths{\sigma^2}$ accounts for the goodness of fit and depends on an unknown variance estimator $\sigma^2$ of the fit residues. To minimize notation overhead, in the following we will refer to estimators of any quantity (\eg~ variance, standard deviation, mean) simply with a letter. $P\pths{K, M}$ can be regarded as a penalty which depends on the sample size $K$ and the number of parameters $M$. In a likelihood framework, the goodness of fit can be expressed in terms of a negative likelihood function \cite{burnham02},
\begin{align}
G\pths{\sigma^2} = - 2\ln\pths{\mathcal{L}(\hattheta)} \ ,
\label{eq:kullback_leibler_info}
\end{align}
where $\mathcal{L}(\hattheta)$ is the likelihood of the candidate model given the data, evaluated at the maximum likelihood estimate of the model parameters $\hattheta$. Computing $\Delta G$ between two models is a measure known as the Kullback-Leibler information \cite{burnham02}. In isolation, Eqs. \ref{eq:information_criteria} and \ref{eq:kullback_leibler_info} are meaningless; only differences in their values when calculated using different models are useful. Different choices of $P\pths{K, M}$ have been proposed, being the Akaike Information Criteria (AIC) \cite{akaike74} and the Bayesian Information Criteria (BIC) \cite{schwarz78} the most common ones:
%
%
\begin{align}
P_\mathrm{AIC} &= 2M + \frac{2M\pths{M+1}}{K-M-1} = \frac{2MK}{K-M-1} \ , \label{eq:aic_penalty} \\
P_\mathrm{BIC} &= M \ln\pths{K} \ , \label{eq:bic_penalty}
\end{align}
where $M$ is the total number of estimated parameters in the model and $K$ is the sample size. The last term in Eq. \ref{eq:aic_penalty} is a correction introduced to compensate the AIC tendency to overfit for finite samples \cite{hurvich89}. BIC has a strong penalty, and therefore can lead to underfitting in large samples \cite{rinke16}. The Weighted Average Information Criteria (WIC) \cite{wu98} provides a solution which performs well, independent of the sample size:
\begin{align}
P_\mathrm{WIC} &= \frac{P_\mathrm{AIC}^2 + P_\mathrm{BIC}^2}{P_\mathrm{AIC} + P_\mathrm{BIC}} \ .
\label{eq:wic}
\end{align}
We use $P_\mathrm{WIC}$ as the the penalty term $P(K,M)$ in Eq. (8) of the main manuscript.


In the limit of large sample sizes $K\gg1$, and assuming normally-distributed residues (errors) with a constant variance, the negative log-likelihood in Eq. \ref{eq:kullback_leibler_info} becomes (up to a constant term) \cite{akaike74}:
\begin{align}
- 2\ln\pths{\mathcal{L}(\hattheta)} \approx K\ln\pths{\sigma^2} \ ,
\label{eq:neg_log_likelihood_variance}
\end{align}
where $\sigma^2$ is the variance estimator for the true variance $\sigma_0^2$ of the residues. Note that the variance estimator is itself a function which depends on the data and the model parameters. For least-squares estimation, the maximum-likelihood estimator for the variance $\sigma^2 = \Sigma / K$, where $\Sigma$ is the residual sum of squares (RSS), is usually used. However, this is a biased estimator. Instead, using the sample variance:
\begin{align}
s^2 = \frac{\Sigma}{K-M-1} = \frac{\sum_{k=1}^K\epsilon_k^2}{K-M-1} \ ,
\label{eq:unbiased_variance_estimator_ic}
\end{align}
where $\epsilon_k$ are the individual residues, provides an unbiased estimator which contributes an additional penalty term and further avoids overfitting \cite{mcquarrie97, rinke16}.

We now derive Eq. \ref{eq:neg_log_likelihood_variance} for our specific FID models.
Assume a random variable $x$ whose probability distribution $f_0(x;\hattheta)$ is defined on the parameters $\hattheta = \cbracks{\theta_m}$, where $m=1\dots M$ and $M$ is the number of fit parameters.
A set of observations $\bvec{x} = \cbracks{x_k}$, where ${k=1\dots K}$, can be regarded as $K$ samples of the distribution $f_0(x;\hattheta)$.
In our experiment, $\hattheta$ are the three hyperfine parameters for each spin plus the three global parameters $p_0$, $\Tn$ and $\tread$ (see main manuscript), and $\bvec{x}$ is the experimental FID trace.
Assuming a model $f(\bvec{x};\hattheta)$ to be a good approximation of the true distribution $f_0$, we want to find estimates for the model parameters $\hattheta$ which most likely produce the measured data record $\bvec{x}$. The likelihood function, which is to be maximized, is a function on the parameters $\hattheta$
\begin{align}
\mathcal{L}(\hattheta) = f(\bvec{x};\hattheta) \ ,
\end{align}
and corresponds to the model $f$ evaluated at the data points $\bvec{x}$, where $\hattheta$ are now parameters to be estimated. Thus, to define a likelihood function we need to make some assumption about the distribution of the data.

We now assume that the observations $\bvec{x}$ are independent and their noise normally distributed. Thus, the approximating model becomes:
\begin{align}
f(\bvec{x};\hattheta)
&= f(x_1,x_2,\ldots,x_K;\hattheta)
= f(x_1;\hattheta) f(x_2;\hattheta) \dots f(x_K;\hattheta) \nonumber\\
&= \pths{\frac{1}{2\pi\sigma^2}}^{K/2} \exp\pths{ -\sum_{k=1}^K \frac{\epsilon_k^2}{2\sigma^2} }  \ ,
\end{align}
%
where $\epsilon_k$ are the residuals from the fitted model and $\sigma^2$ is an estimator of their variance. It is useful to work with logarithms because they are continuous and monotonous, thus the negative log-likelihood becomes:
\begin{align}
-2\ln\pths{ \mathcal{L}(\hattheta) } &= K\ln\pths{2\pi\sigma^2} + \frac{\sum_{k=1}^K\epsilon_k^2}{\sigma^2} \approx K\ln\pths{2\pi\sigma^2} + K \label{eq:supp_wm_spec_loglikelihood_1}\\
-2\ln\pths{ \mathcal{L}(\hattheta) } &\approx K\ln\pths{\sigma^2} \ ,
\label{eq:supp_wm_spec_loglikelihood_2}
\end{align}
where the approximation holds for large $K$ and we have dropped the term $K\pths{1 + \ln\pths{2\pi}}$ because it is a constant offset \cite{akaike74, hurvich89}.
In our case, the data record $\bvec{x}$ consists of a signal and a noise contribution.  The latter is dominated by photon shot noise, which in the limit of many collected photons converges to a normal distribution. Using Eq. \ref{eq:unbiased_variance_estimator_ic} as variance estimator we thus write:
\begin{align}
K\ln\pths{\sigma^2} = K\ln\pths{\frac{\sum_{k=1}^K\pths{x_k - \tilde{x}_k(\hattheta)}^2}{K-M-1} } \ ,
\label{eq:supp_aic_ls_sim}
\end{align}
where the function $\tilde{\bvec{x}}(\hattheta)$ represents the chosen FID model for the set of parameters $\hattheta$.  
Eq. (\ref{eq:supp_aic_ls_sim}) represents the $G(\hattheta,\bvec{x})$ term in Eqs. (8,9) in the main manuscript.

In our analysis, we fit the complex Fourier spectra instead of the time-domain FID traces.  
Denoting spectral quantities with a hat, we have for each complex spectral component
%
\begin{align}
\hat{x}_j = \realpartfft_j + i \imagpartfft_j = \sum_{k=1}^{K}x_k e^{-i2\pi k j} \ .
\end{align}
The noise of the Fourier spectra is also normally distributed \cite{boss17}.
Since we assume the measurement points $\bvec{x}$ to be independent, the noise in each the real $\realpartfft_j$ and imaginary $\imagpartfft_j$ parts of the complex spectrum also follows a normal distribution, that is, $\realpartfft_j, \imagpartfft_j \sim N\pths{\mu=0, \sigma_0^2}$ for $K\gg1$. $\realpartfft_j$ and $\imagpartfft_j$ are independent for $j < K/2$ \cite{boss17}.

In terms of the power spectrum, its components are $\hat{y}_j = |\hat{x}_j|^2 = \realpartfft_j^2 + \imagpartfft_j^2$, meaning that their noise follows a chi-squared distribution with two degrees of freedom ($\chi^2_2$). Therefore, the residues from the fitted model $\hat{\epsilon}_j = \hat{y}_j - \hat{\tilde{y}}_j(\hattheta)$ would follow a displaced chi-squared distribution $\chi^2_2\pths{\frac{\hat{\epsilon}_j + \mu}{\sigma}} = \chi^2_2\pths{\frac{\hat{\epsilon}_j + \sigma}{\sigma}}$, where we use standardized units and the fact that the expectation value $\mu$ for a $\chi^2_2$ distribution equals its standard deviation $\sigma$. The model becomes:
\begin{align*}
f\pths{\hat{\bvec{y}}, \hattheta} &= f\pths{\hat{y}_1, \hattheta}f\pths{\hat{y}_2, \hattheta}\ldots f\pths{\hat{y}_K, \hattheta} \\
&= \chi^2_2\pths{\frac{\hat{\epsilon}_1 + \mu}{\sigma}}\chi^2_2\pths{\frac{\hat{\epsilon}_2 + \mu}{\sigma}}\ldots\chi^2_2\pths{\frac{\hat{\epsilon}_K + \mu}{\sigma}} \\
&= \pths{\frac{1}{2\sigma}}^K\exp\pths{ -\frac{1}{2}\sum^K_{j=1} \frac{\hat{\epsilon}_j + \sigma}{\sigma} } = \pths{\frac{1}{2\sigma}}^K e^{-K/2} \ ,
\end{align*}
where we have taken into account that $\sum^K_{j=1} \hat{\epsilon}_j$ follows a normal distribution with zero mean (central-limit theorem for $K\gg 1$). Note the normalization factor $1/\sigma$ multiplying the $\chi^2_2$ distributions such that their integrals are still unity. Taking the negative logarithm we find:
\begin{align*}
-2\ln\pths{\mathcal{L}(\hattheta)} &= -2\ln\pths{f\pths{\hat{\bvec{y}}, \hattheta}} = 2K\ln\pths{2\sigma} + K \nonumber\\ &\approx K\ln\pths{\sigma^2} \ ,
\end{align*}
where we express the standard deviation as the square root of the variance estimator $\sigma = \sqrt{\sigma^2}$ and again dropped constant offsets. Using again Eq. \ref{eq:unbiased_variance_estimator_ic}, the negative log-likelihood function for the power spectrum becomes:
\begin{align}
-2\ln\pths{\mathcal{L}(\hattheta)} \approx K\ln\pths{\sigma^2} = K\ln\pths{ \frac{\sum^K_{j=1} \pths{\hat{y}_j - \hat{\tilde{y}}_j\pths{\hattheta}}^2}{K-M-1} } \ ,
    \label{eq:nll_psd}
\end{align}
where $\hat{\tilde{\bvec{y}}}\pths{\hattheta} = \{\hat{\tilde{y}}_j\}_{j=1\cdots K}$ represents the power spectrum of the chosen FID model $\tilde{\bvec{x}}(\hattheta)$.
%

For a number $D$ of independent datasets (\ie~FID time traces), we then write the combined likelihood function for the spectral components and their square magnitude as:
\begin{align}
- 2\ln\pths{\mathcal{L}(\hattheta)} &= - 2\ln\pths{\prod_{d=1}^D\mathcal{L}_d(\hattheta)} = - 2\ln\pths{\prod_{d=1}^D \mathcal{L}_{d,re}(\hattheta)\mathcal{L}_{d,im}(\hattheta)\mathcal{L}_{d,psd}(\hattheta) } \nonumber\\
&= -2\sum_{d=1}^D \ln\pths{\mathcal{L}_{d,re}(\hattheta)\mathcal{L}_{d,im}(\hattheta)\mathcal{L}_{d,psd}(\hattheta)} \nonumber\\
&= - 2\sum_{d=1}^D\pths{ \ln\pths{\mathcal{L}_{d,re}(\hattheta)} + \ln\pths{\mathcal{L}_{d,im}(\hattheta)} + \ln\pths{\mathcal{L}_{d,psd}(\hattheta)} } \nonumber\\
&= K\sum_{d=1}^D \pths{\ln\pths{\sigma^2_{d,re}} + \ln\pths{\sigma^2_{d,im}} + \ln\pths{\sigma^2_{d,psd}}} \ ,
\label{eq:joint_aic}
\end{align}
where the subscripts $(\cdot)_{re}$, $(\cdot)_{im}$ and $(\cdot)_{psd}$ label the real, imaginary and magnitude-squared parts of the power spectra. Using Eq. \ref{eq:unbiased_variance_estimator_ic} for the variance estimators, we finally write:
\begin{align}
- 2\ln\pths{\mathcal{L}(\hattheta)} = - 3DK\ln\pths{K-M-1} + K\sum_{d=1}^D \sum_{ic} \ln\pths{\Sigma_{d,ic}} \ ,
\label{eq:joint_aic_rss}
\end{align}
where $\Sigma_{d,ic}$ is the residual sum of squares for dataset $d$ and $ic\in\cbracks{re, im, psd}$.
To find maximum likelihood estimates using the WIC we thus add Eq. \ref{eq:joint_aic_rss} and Eq. \ref{eq:wic}.

\subsection{Generalized Simulated Annealing (GSA)}
GSA is a stochastic approach \cite{tsallis96} that combines the original method of Classical Simulated Annealing (CSA) \cite{kirkpatrick83} and Fast Simulated Annealing (FSA) \cite{szu87}. It has proven very useful to find global minima of complicated, multidimensional systems with large numbers of local minima such those found in quantum chemistry, genetics or non-linear time-series. Due to its statistical nature, local minima can be escaped much more easily than with steepest-descent or gradient methods \cite{xiang97}. The core idea is to combine importance sampling with an artificial temperature which is gradually decreased to simulate thermal noise.
The algorithm uses a distorted Cauchy-Lorentz visiting distribution, to sample the parameter space \cite{tsallis96}:
\begin{align}
g_{q_v}\pths{\Delta x_t} \propto \frac{T_{q_v}\pths{t}^{-D/\pths{3-q_v}}}{\bracks{1+\pths{q_v-1} \frac{\pths{\Delta x_t}^2}{\bracks{T_{q_v}\pths{t}}^{2/\pths{3-q_v}} } }^{1/\pths{q_v-1}+\pths{D-1}/2} } \ ,
\label{eq:gsa_distribution}
\end{align}
where $D$ is the number of dimensions (\ie ~fit parameters). Eq. \ref{eq:gsa_distribution} is also known as a Tsallis distribution. Compared to the Boltzmann (CSA) or Cauchy (FSA) distributions, Eq. \ref{eq:gsa_distribution} has a longer tail and thereby combines a local search with frequent long jumps which facilitate detrapping from local minima. The shape of the distribution is controlled by the visiting parameter $q_v$, which also controls the cooling rate \cite{tsallis96}:
\begin{align}
T_{q_v}\pths{t} = T_{q_v}\pths{1}\frac{2^{q_v -1}-1}{\pths{1+t}^{q_v -1} -1} \ .
\label{eq:gsa_temperature}
\end{align}
At (computational) time $t$ and artificial visiting temperature $T_{q_v}\pths{t}$, GSA thus requires the generation of random jumps $\Delta x_t$ distributed according to Eq. \ref{eq:gsa_distribution}. Here, we use an exact D-dimensional Tsallis random number generator \cite{schanze06}. A new configuration $x_t =  x_{t-1}+\Delta x\pths{t}$ is accepted with a probability \cite{tsallis96}:
\begin{align}
p_{q_a} = \min\cbracks{1, \bracks{1-\pths{1-q_a}\beta\Delta E}^{1/\pths{1-q_a}}} \ ,
\label{eq:gsa_acceptance_probability}
\end{align}
where $q_a$ is an acceptance parameter and $\beta\propto T_a\pths{t}^{-1}$ is inverse acceptance temperature. $T_a\pths{t}$ is also controlled by $q_v$ according to the cooling rate given by Eq. \ref{eq:gsa_temperature}. Eq. \ref{eq:gsa_acceptance_probability} is a generalized Metropolis algorithm. Configurations with a lower cost ($\Delta E<0$) are thus always accepted. For $q_a < 0$, zero acceptance probability is assigned to the cases where \cite{tsallis96}:
\begin{align}
1-\pths{1-q_a}\beta\Delta E < 0 \ ,
\label{eq:gsa_acceptance_probability_qa}
\end{align}
meaning that $\Delta E$ must be smaller than $\beta/\pths{1-q_a}$ for hill-climbing to occur. The choice of $q_a<0$ in the argument of the power-law acceptance function (Eq. \ref{eq:gsa_acceptance_probability}) was originally suggested because it favors lower energies. However, alternative choices of $p_{q_a}$ for $q_a>1$ have also been proposed \cite{andricioaei96}. Setting $q_v=q_a=1$ recovers CSA, and $q_v=2$, $q_a=1$ recovers FSA \cite{tsallis96}. As $T\rightarrow 0$ (or $t\rightarrow \infty$), the algorithm becomes a steepest-descent method \cite{xiang97}.

For our calculations, we selected $q_v=2.7$, $q_a=-1$ and a large initial acceptance temperature $T_{a}\pths{1}\sim 10^3$ to facilitate the escape from local minima. 
A technique to accelerate convergence is to set:
\begin{align}
T_a\pths{t} \rightarrow \frac{T_a\pths{t}}{t} \ ,
\end{align}
which is equivalent to a monotonously decreasing $q_a \propto t^{-1}$ \cite{xiang97}. It is worth noting that the computational times associated with the visiting and acceptance temperatures need not to be equal. To improve convergence towards physically-relevant configurations, we modified the acceptance probability $p_{q_a} \rightarrow p_{q_a}'$ using a step-function:
\begin{align}
p_{q_a}' &= p_{s} \cdot p_{q_a} \label{eq:gsa_acceptance_probability_mod} \\
p_{s} &= \begin{cases}
1 & \text{if } d_{ij} > r_{cc} \; \forall i\neq j \\
0 & \text{otherwise}
\end{cases} \quad ,
\label{eq:gsa_p_sens}
\end{align}
where $d_{ij}$ is the estimated spatial distance between any two spins $i$ and $j$ at any computational time $t$, and $r_{cc} = 0.154\unit{nm}$ is the diamond bond-length. In other words, we reject any proposed nuclear-spin configurations where any two spins are closer to each other than the smallest \C-\C distance in a diamond crystal.

We also assigned an individual visiting temperature $T_{q_v, m}$ to each fit parameter $\theta_m$ in order to account for their different scales/units, and set their initial values large enough such that their entire search intervals were covered at $t=0$ (see Sec. \ref{sec:parameter_intervals}). Furthermore, we observed better results when normalizing the visiting and acceptance times to a few hundred Monte Carlo steps. Specifically, we used $t_v = t/400$ and $t_a=t/200$. This re-scaling of the computational time effectively helps to more widely explore the parameter space at each temperature step. We also observed better convergence when updating one parameter at a time instead of the whole parameter set.

For a given number $n$ of spins, we run the algorithm starting from $\sim 10^2-10^3$ initial random spin configurations, where each configuration is drawn by randomly selecting $n$ points within a radius of $2.5\unit{nm}$ around the coordinate origin. We ran the algorithm over 5000 iterations and took the best result. As a guide, a single run (1 initial spin configuration) of the algorithm for $n\approx 20$ took $\sim 3\unit{h}$.

It is worth noting that the performance of GSA, and in general of annealing algorithms, is problem dependent and therefore some parameter tuning is usually required. Our selections for $q_v$ and $q_a$ lie within ranges where good performance has been reported \cite{tsallis96}.

\subsection{Estimation of uncertainties}
As mentioned in the main text, we simultaneously fitted all spectra. To get an estimate for the uncertainties in the fit parameters $\hattheta$, we resampled the data using a bootstrapping protocol \cite{efron79}.  We calculated the mean and standard deviation of the fit residues (green dots in Fig. 4b,c in the main text) for the $\ReFFT$ and $\ImFFT$ parts of each spectrum.  We then used the resulting deviations to generate normally-distributed noise and thereby created a set of $P=100$ new, artificial datasets. Each newly generated dataset (set of 4 spectra) was minimized again (using a steepest-descent search) and the original-dataset solution as starting point. We thereby generated a set $\theta_{m,p} \equiv \cbracks{\theta_m}_p$ of estimates for each fit parameter $\theta_m$, where $p=1\dots P$.  Finally, we calculated the fit results as:
\begin{align}
\begin{split}
\bar\theta_m &= \exvalue{\theta_{m,p}} = \frac{1}{P}\sum_{p=1}^P \theta_{m,p} \\
\sigma(\theta_m) &= \pths{\frac{1}{P-1}\sum_{p=1}^P \pths{\theta_{m,p} - \bar\theta_m}^2}^{1/2}
\end{split}
\label{eq:wm_aic_fit_results}
\end{align}
%

\subsection{Selection of parameter intervals}\label{sec:parameter_intervals}

To start our minimization routines, we chose random starting values from a pre-defined interval for each parameter. 
%

For $\apari$, we defined the interval by looking at the spectral support across all power spectra and set it to $2\pi \times [2\pths{f_0-f_k} \dots 2\pths{f_0+f_k}]$, where $f_0$ is the central-peak frequency (assumed to be around the bare Larmor frequency) and we selected $f_k$ such that all visible peaks in the power spectra where covered. Specifically, we selected $f_k= 4.5\unit{kHz}$ and $f_k= 4\unit{kHz}$ for dataset 1 and 2, respectively.

%
For $\aperpi$, we defined a sufficiently large interval to cover the minimum and maximum observable couplings. Specifically, we chose $2\pi \times [0.0 \dots 80\unit{kHz}]$ and $2\pi \times [0.0 \dots 70\unit{kHz}]$ for dataset 1 and dataset 2, respectively.
%

We initialized $\phii$ in the interval $[0.0 \dots 360\unit{\degree}]$, but constrained it to the interval $[-90.0 \dots 450\unit{\degree}]$ to avoid phase-wrapping with values close to $0\degree$ or $360\degree$.
Finally, we assumed starting intervals for the global parameters as $p_0 \in [0.3 \dots 1.0]$, $\Go\in [0\dots \pths{12\unit{ms}}^{-1}]$ and $\tread\in [0 \dots \ts]$.

\subsection{Global parameters $p_0$, $\Tn$ and $\tread$}

On top of the measurement-induced dephasing with rate $\Gamma_\beta = \beta^2/(4\ts)$, nuclear spins are also subject to intrinsic dephasing with rate $\Gamma_0 = (\Tn)^{-1}$ due to $T_2^\ast$ relaxation and further relaxation with rate $\Gammal$ caused by the optical readout.
Since $T_2^\ast$ dephasing in our experiment is dominated by fluctuations in the bias field, we can treat $\Gamma_0$ as a global fit parameter.
The readout-induced dephasing is explained as follows: owing to our off-resonant optical readout and re-initialization scheme, the NV center undergoes stochastic spin-state, electronic-state and potentially charge-state transitions during optical illumination before reaching the polarized $m_S=0$ state \cite{waldherr11,dhomkar16}.
The residual hyperfine interaction during laser readout causes an additional dephasing with rate \cite{cujia19}:
\begin{align*}
\Gammal = \frac{\pths{\apar \tread}^2}{2\ts}
\end{align*}
where $\tread$ is a phenomenological time constant that is roughly of the duration of the laser pulse.  $\tread$ is not correlated with the \C environment \cite{waldherr11,dhomkar16} and is a global free fit parameter in our maximum likelihood estimation.
The total decay rate is then the sum of all contributions
\begin{align}
\Gamma = \Gamma_\beta + \Gamma_0 + \Gammal = \frac{\beta^2}{4\ts} + \frac{1}{\Tn} + \frac{\apar^2\tread^2}{2\ts}
\label{eq:supp_total_decay_global}
\end{align}
The third parameter, the initial nuclear polarization $p_0$, in principle differs for every \C nuclei, because the polarization rate is proportional to $\aperp$.  However, because we repeat the polarization transfer for typically $>10^3$ cycles, we expect that the polarization of all nuclei within a resonant slice become saturated.  This assumption is supported by the fact that spectra show little change in peak intensities once the number of cycles exceeds $10^3$ (see Ref. \onlinecite{cujia19}, Extended Data Fig. 5) and is consistent with similar models using cross-relaxation induced polarization (CRIP) \cite{broadway18}.  Therefore, we assume $p_0$ to become homogeneous within our sensitive slices and thus treat it as a global parameter.
%
%
\section{Weak measurement spectroscopy}

\subsection{Single-spin free induction decay signal under continuous weak measurements}
 
Weak measurements probe the transverse nuclear polarization (specifically, the $\exvalue{\Ix}$ component) linearly with the measurement strength $\beta$ \cite{cujia19}. The measured free-induction decay (FID) signal for nuclear spin $i$ as a function of time, that is, the measured transition probability for the spin sensor, has the form:
\begin{align}
x_i\pths{t} = \frac{1}{2}p_0\sinf{\beta}\cosf{\omega_{i} t}e^{-\Gamma_i t} e^{-\pths{\tbeta/\Te}^a}
\label{eq:supp_single_fid}
\end{align}
where $\omega_{i} = \magn{\gamma_n \int_0^{t} B\pths{t'}dt'} $ is the average precession frequency of nuclear spin $i$. For completeness, we have included the NV electronic spin coherence time $\Te$, where $a$ is an exponent which describes the dephasing rate of the spin sensor (exponential or Gaussian).  Because in our experiments $\tbeta\ll\Te$, sensor dephasing is not important for our study. $p_0$ describes the initial nuclear polarization. 

Weak measurements simultaneously address all nuclear spins within the detector bandwidth of approximately $\tbeta^{-1}$ \cite{cujia19}. Because the measurements are weak, $\beta\approx\aperp\tbeta\ll 1$, we can neglect nuclear-nuclear interactions mediated by the sensor spin. For $n$ nuclear spins, the total FID signal is thus the superposition of all individual contributions, leading to Eq. (2) in the main text.

\section{Sensitive slice}\label{sec:s_sensitive_slice}
We define the power signal-to-noise ratio (pSNR) as the ratio between the spectral peak amplitude and the standard deviation in a reference spectral region where there are no signals. For a single nuclear spin, the total signal after $K$ weak measurements is given by \cite{cujia19}:
\begin{align}
S=\frac{\epsilon C p_0 \sinf{\beta}}{2\Gamma \ts}\pths{1-e^{-\Gamma K\ts}} \ ,
\label{eq:snr_signal_amplitude}
\end{align}
where $C$ represents the total photon counts, $\epsilon$ the optical contrast, and $p_0\cbr{0...1}$ is the nuclear polarization.
Two processes contribute to the total noise in a single weak measurement, a Bernoulli process resulting from quantum projection noise and a Poisson process associated with photon shot noise \cite{boss17}.  In our experiments, the photon shot noise greatly exceeds the quantum projection noise, and the noise variance is given by:
\begin{align}
\sigma_x^2 = C\pths{1-\frac{\epsilon}{2}}\approx C 
\label{eq:snr_noise_variance}
\end{align}
where the last approximation is valid for small $\epsilon\ll 1$, which is usually the case in our experiments. The power SNR is given by:
\begin{align}
\pSNR = \frac{S^2}{K\sigma_x^2} \approx \pths{\frac{\epsilon C p_0\sinf{\beta}}{2\Gamma \ts}\pths{1-e^{-\Gamma K\ts}}}^2\frac{1}{KC} \ .
\label{eq:power_snr}
\end{align}
To obtain the amplitude SNR per unit time, we divide by the duration $t_\mr{m} = \tpol + K\ts$ of one complete measurement cycle and take the square root:
\begin{align}
\SNR_0 = \sqrt{\frac{\pSNR}{t_\mr{m}}} \approx \frac{\epsilon \sqrt{C}p_0\sinf{\beta}}{2\Gamma\sqrt{K} \ts}\frac{\pths{1-e^{-\Gamma K\ts}}}{\sqrt{t_\mr{m}}} \ .
\label{eq:snr_per_time}
\end{align}
We now define the sensitivity function $\mathcal{S}\pths{\beta}$ by the amplitude SNR expressed as a function of $\beta$, dropping the count rate factor $\epsilon\sqrt{C}$ and taking the approximation $\sin(\beta) \approx \beta$:
\begin{align}
\mathcal{S}\pths{\beta} = \frac{\SNR_0}{\epsilon\sqrt{C}} \overset{\beta\ll 1}{\approx} \frac{p_0\beta}{2\Gamma\sqrt{K} \ts}\frac{\pths{1-e^{-\Gamma K\ts}}}{\sqrt{t_\mr{m}}} \ .
\label{eq:sensitivity_function}
\end{align}
This equation corresponds to Eq. (6) in the main manuscript.  Eq. \ref{eq:sensitivity_function} is maximized in the regions of large $\aperp$ and vanishing $\apar$, and reaches an optimum point \cite{cujia19} when the intrinsic and induced decay rates are commensurate, \ie, $\Gamma_0 \approx \Gamma_{\beta}$ and thus $\Gamma \approx 2\Gamma_\beta = \beta^2/\pths{2\ts}$.
To maximize signal acquisition while minimizing decay, we set the total measurement time to $K\ts = 1/\Gamma$, meaning that:
\begin{align}
\beta^{\pths{\mathrm{opt}}} = \sqrt{\frac{2}{K}} \ .
\label{eq:optimal_beta}
\end{align}
Plugging in these conditions into Eq. \ref{eq:sensitivity_function}, assuming comparable initialization and measurement times $\tpol \approx K\ts$ and taking into account that $\ts \approx \tbeta$ (short sensor readout times) we obtain:
\begin{align}
\mathcal{S}(\beta^{\pths{\mathrm{opt}}}) \propto \frac{\aperp\sqrt{\tbeta}}{2\sqrt{2}\pi}
\label{eq:sens_func_aperp}
\end{align}
Since $\aperp \propto r^{-3}$, the sensitivity radius scales as $r \propto \tbeta^{1/6}$.
%
%
\section{Internuclear couplings}
Continuous weak measurements can be regarded as the equivalent of inductive detection in conventional NMR. In this way, they capture the full evolution of the nuclear environment, including potential nuclear spin-spin interactions.  Such nuclear spin-spin interactions lead to additional peaks in the spectra, which can interfere with our minimization strategy.  To illustrate the effect of nuclear spin-spin interactions and analyze whether they affect our present experiment, we performed a density matrix simulation with $n=3$ randomly positioned nuclear spins using a Master equation approach.

Fig. \ref{fig:nuclear_couplings} shows the simulated weak measurement spectrum for nuclear spins with inter-nuclear coupling constants $g_{12}=634.79\unit{Hz}$, $g_{13}=1.17\unit{Hz}$, and $g_{23}=22.05\unit{Hz}$.  The coupling constants are chosen arbitrarily.  All splittings (orange arrows) within the FFT resolution ($\ts=8\unit{\us}$) are visible, illustrating how weak measurement-spectroscopy gives access to the whole dynamics. The direct consequence is that weak measurement spectroscopy can potentially yield very complex spectra, similar to inductively-detected Fourier NMR spectroscopy.



To estimate whether such couplings are potentially present in our experimental spectra, we computed the average number of spin pairs yielding resolvable couplings within our spectral resolution given by  $\Gamma=1/\min\{K\ts,\pi \Tn\}$. The FFT resolution is $156\unit{Hz}$ and $110\unit{Hz}$ for \NVa and \NVbns, and the estimated nuclear linewidth is $1/(\pi \Tn)\approx 300\ldots400\unit{Hz}$. We generate $10^4$ random \C nuclear spin configurations within the sensitive volumes of \NVa and \NVb (Fig. 5a,d in the main text), respectively, and count the number of pairs with couplings larger than $\Gamma$.  On average, we find $1.14\pm 1.06$ ($69\%$ probability that at least one spin pair is present) and $2.14\pm 1.43$ ($89\%$ probability) spin-pairs for \NVa and \NVb, respectively. These figures reduce to $0.28\pm 0.52$ ($25\%$ probability) and $0.43\pm 0.64$ ($36\%$ probability) for the estimated NMR resolution $>300\unit{Hz}$.
Thus, although unlikely, we can not rule out the presence of an unidentified spin-pair in the spectra. 
As a control, we have inspected the reconstructed spatial positions of the mapped nuclei and computed their relative spatial distances.  We find an average distance between the closest pairs of $0.212\unit{nm}$ and $0.178\unit{nm}$ for dataset 1 and 2, respectively. This corresponds to homonuclear couplings of $544\unit{Hz}$ and $645\unit{Hz}$.  However, due to the inversion symmetry of the diamond crystal, each recovered nuclear position has a 50\% probability to be mirrored along the coordinate origin.




\clearpage
\section{Supplementary Tables}

\begin{table}[h!]
    \begin{tabular}{l|l}
        \textit{Figure 4b} & \\
        NV center & $1$ \\
        B field & 201.29$\unit{mT}$\\
        Electronic $\Te$ (CPMG) & $216\unit{\mu s}$\\
        NV initialization laser pulse & $1.5\unit{\mu s}$\\
        NV readout laser pulse & $1.5\unit{\mu s}$\\
        Weak measurement CPMG pulses $K$ & $\{4,8,12,16\}$\\
        Weak measurement CPMG duration $\tbeta$ & $\cbracks{0.930, 1.860, 2.790, 3.720} \unit{\mu s}$\\
        Number of weak measurements $n$ & $800$\\
        Sampling period $\ts$ & $8.0 \unit{\mu s}$\\
        Duration of LR-NOVEL pulse & $30\unit{\us}$\\
        Ramp amplitude of LR-NOVEL pulse & $10\unit{\%}$\\
        $\pi/2$ rotation RF pulse & $10.903\unit{\us}$\\
        Number of polarization repetitions $M$ & $1200$ \\
        Total integration time & $10\unit{h}$, equal to $2.5\unit{h}$ for each $\tbeta$
    \end{tabular}\hfill\
		\caption{Measurement parameters for Figure 4b.  A detailed pulse diagram is given the Extended Data Fig. 1b of Ref. \onlinecite{cujia19}.}
    \label{tbl:params_fig2a}
\end{table}

\begin{table}[h!]
    \begin{tabular}{l|l}
        \textit{Figure 4c} & \\
        NV center & $2$ \\
        B field & $188.89\unit{mT}$\\
        Electronic $\Te$ (CPMG) & $80\unit{\mu s}$\\
        NV initialization laser pulse & $1.5\unit{\mu s}$\\
        NV readout laser pulse & $1.5\unit{\mu s}$\\
        Weak measurement CPMG pulses $K$ & $\cbr{12,16,20,24}$\\
        Weak measurement CPMG duration $\tbeta$ & $\cbracks{2.966, 3.956, 4.944, 5.932} \unit{\mu s}$\\
        Number of weak measurements $n$ & $800$\\
        Sampling period $\ts$ & $11.48 \unit{\mu s}$\\
        Duration of LR-NOVEL pulse & $25\unit{\us}$\\
        Ramp amplitude of LR-NOVEL pulse & $10\unit{\%}$\\
        $\pi/2$ rotation RF pulse & $20.520\unit{\us}$\\
        Number of polarization repetitions $M$ & $1600$ \\
        Total integration time & $45.73\unit{h}$, equal to $11.43\unit{h}$ for each $\tbeta$
    \end{tabular}\hfill\
		\caption{Measurement parameters for Figure 4c. The parameters for the FID shown in \textit{Fig. 4a} are those for $\tbeta=4.944\unit{\us}$.  A detailed pulse diagram is given the Extended Data Fig. 1b of Ref. \onlinecite{cujia19}.}
    \label{tbl:params_fig2b}
\end{table}
%

\clearpage
\addtolength{\tabcolsep}{2pt}
\begin{table}[h]
\centering
\begin{tabular}{crrrrrr}
    \hline
    \C  & $\apar\pths{\ftow{\unit{kHz}}}$   & $\aperp\pths{\ftow{\unit{kHz}}}$   & $r\pths{\unit{nm}}$   & $\theta\pths{\unit{deg}}$   & $\phi\pths{\unit{deg}}$   & $\delta V$ (\AA$^3$)   \\
    \hline
    1   & $-8.14 \pm 0.12$                  & $36.40 \pm 4.66$                   & $0.89 \pm 0.03$       & $60.88 \pm 0.77$            & $137.20 \pm 6.21$         & $0.26$               \\
    2   & $-5.47 \pm 0.08$                  & $22.80 \pm 3.97$                   & $1.04 \pm 0.05$       & $61.38 \pm 1.16$            & $322.71 \pm 17.27$        & $2.22$               \\
    3   & $-5.10 \pm 0.06$                  & $9.07 \pm 2.76$                    & $1.32 \pm 0.11$       & $68.77 \pm 3.82$            & $312.55 \pm 30.21$        & $29.09^\ast$              \\
    4   & $-3.41 \pm 0.03$                  & $21.53 \pm 2.85$                   & $1.07 \pm 0.04$       & $59.18 \pm 0.59$            & $13.96 \pm 10.49$         & $0.87$               \\
    5   & $-3.18 \pm 0.03$                  & $9.43 \pm 3.02$                    & $1.38 \pm 0.16$       & $64.27 \pm 3.67$            & $59.99 \pm 12.13$         & $20.29^\ast$              \\
    6   & $-1.91 \pm 0.04$                  & $7.54 \pm 2.08$                    & $1.51 \pm 0.14$       & $62.05 \pm 2.46$            & $250.42 \pm 19.13$        & $25.72^\ast$              \\
    7   & $-1.81 \pm 0.03$                  & $16.95 \pm 2.26$                   & $1.17 \pm 0.05$       & $57.77 \pm 0.42$            & $345.77 \pm 13.86$        & $1.58$               \\
    8   & $-0.69 \pm 0.03$                  & $18.40 \pm 2.67$                   & $1.15 \pm 0.06$       & $55.83 \pm 0.17$            & $48.00 \pm 8.45$          & $1.16$               \\
    9   & $-0.69 \pm 0.04$                  & $36.80 \pm 3.03$                   & $0.91 \pm 0.03$       & $55.28 \pm 0.05$            & $118.14 \pm 12.01$        & $0.26$               \\
    10  & $-0.27 \pm 0.02$                  & $17.63 \pm 2.89$                   & $1.17 \pm 0.07$       & $55.18 \pm 0.09$            & $160.57 \pm 7.13$         & $1.39$               \\
    11  & $-0.15 \pm 0.05$                  & $23.36 \pm 3.61$                   & $1.07 \pm 0.06$       & $54.93 \pm 0.07$            & $36.75 \pm 10.71$         & $1.39$               \\
    12  & $0.12 \pm 0.01$                   & $9.83 \pm 2.77$                    & $1.45 \pm 0.19$       & $54.34 \pm 0.33$            & $325.24 \pm 15.78$        & $31.24^\ast$              \\
    13  & $0.46 \pm 0.05$                   & $23.67 \pm 3.17$                   & $1.07 \pm 0.05$       & $54.17 \pm 0.10$            & $137.58 \pm 9.80$         & $0.94$               \\
    14  & $1.02 \pm 0.03$                   & $12.03 \pm 2.12$                   & $1.35 \pm 0.09$       & $52.19 \pm 0.52$            & $262.02 \pm 12.74$        & $4.83$               \\
    15  & $1.78 \pm 0.05$                   & $3.20 \pm 0.82$                    & $2.11 \pm 0.16$       & $38.11 \pm 4.31$            & $312.18 \pm 25.23$        & $101.10^\ast$             \\
    16  & $2.76 \pm 0.13$                   & $75.59 \pm 3.41$                   & $0.72 \pm 0.01$       & $53.68 \pm 0.07$            & $179.94 \pm 12.90$        & $0.04$               \\
    17  & $2.81 \pm 0.08$                   & $59.09 \pm 3.50$                   & $0.79 \pm 0.02$       & $53.36 \pm 0.09$            & $130.90 \pm 9.42$         & $0.07$               \\
    18  & $3.53 \pm 0.07$                   & $30.62 \pm 3.58$                   & $0.99 \pm 0.04$       & $51.34 \pm 0.42$            & $229.59 \pm 13.11$        & $0.80$               \\
    19  & $4.56 \pm 0.08$                   & $8.24 \pm 4.01$                    & $1.49 \pm 0.49$       & $37.57 \pm 10.99$           & $251.81 \pm 40.95$        & $710.58^\ast$             \\
    20  & $4.83 \pm 0.09$                   & $8.18 \pm 3.47$                    & $1.52 \pm 0.35$       & $36.46 \pm 9.19$            & $290.20 \pm 40.89$        & $409.71^\ast$             \\
    \hline
\end{tabular}
\caption{\C nuclear spins localization results for NV 1.  Fit results for global parameters are $p_0=0.66\pm0.03$, $\Tn=11.69\pm 1.03\unit{ms}$ and $\tread=1.65\pm 0.25\unit{\us}$. Errors for $\apar$, $\aperp$ and $\phi$ are calculated by bootstrapping. Errors for $r$, $\vartheta$, $\phi$ and the global parameters are calculated using Monte Carlo error propagation. The estimated uncertainty volume $\delta V$ (also depicted in Fig. 5c) is given in cubic Angstrom.  $^\ast$ denote spins whose uncertainty volume is larger than the volume per carbon atom in the diamond lattice (5.69\,\AA$^3$).
}
\label{tbl:wm_loc_results_1}
\end{table}

\clearpage
\begin{table}[h!]
    \centering
    \begin{tabular}{crrrrrr}
        \hline
        \C  & $\apar\pths{\ftow{\unit{kHz}}}$   & $\aperp\pths{\ftow{\unit{kHz}}}$   & $r\pths{\unit{nm}}$   & $\theta\pths{\unit{deg}}$   & $\phi\pths{\unit{deg}}$   & $\delta V$ (\AA$^3$)   \\
        \hline
        1   & $-5.60 \pm 0.02$                  & $4.06 \pm 0.60$                    & $1.45 \pm 0.02$       & $77.72 \pm 1.47$            & $164.47 \pm 8.24$         & $0.74$               \\
        2   & $-4.78 \pm 0.20$                  & $57.29 \pm 4.17$                   & $0.78 \pm 0.02$       & $57.10 \pm 0.20$            & $163.35 \pm 6.66$         & $0.07$               \\
        3   & $-3.92 \pm 0.08$                  & $39.69 \pm 2.56$                   & $0.88 \pm 0.02$       & $57.52 \pm 0.19$            & $289.33 \pm 5.14$         & $0.06$               \\
        4   & $-1.50 \pm 0.04$                  & $1.55 \pm 0.32$                    & $2.17 \pm 0.06$       & $74.12 \pm 2.29$            & $310.91 \pm 31.43$        & $27.20^\ast$              \\
        5   & $-1.26 \pm 0.01$                  & $8.34 \pm 1.10$                    & $1.47 \pm 0.06$       & $58.98 \pm 0.56$            & $188.47 \pm 5.48$         & $1.18$               \\
        6   & $-0.68 \pm 0.17$                  & $49.71 \pm 5.63$                   & $0.83 \pm 0.03$       & $55.13 \pm 0.11$            & $99.38 \pm 6.51$          & $0.20$               \\
        7   & $-0.38 \pm 0.02$                  & $24.89 \pm 1.74$                   & $1.04 \pm 0.02$       & $55.17 \pm 0.04$            & $207.35 \pm 5.99$         & $0.14$               \\
        8   & $-0.17 \pm 0.01$                  & $9.33 \pm 1.50$                    & $1.45 \pm 0.08$       & $55.28 \pm 0.10$            & $320.92 \pm 6.50$         & $2.27$               \\
        9   & $-0.10 \pm 0.01$                  & $8.97 \pm 1.62$                    & $1.47 \pm 0.09$       & $55.08 \pm 0.08$            & $292.63 \pm 6.46$         & $3.09$               \\
        10  & $-0.03 \pm 0.01$                  & $22.18 \pm 2.10$                   & $1.08 \pm 0.03$       & $54.77 \pm 0.01$            & $312.39 \pm 3.27$         & $0.16$               \\
        11  & $0.14 \pm 0.01$                   & $9.66 \pm 1.24$                    & $1.44 \pm 0.06$       & $54.31 \pm 0.06$            & $112.11 \pm 8.04$         & $1.72$               \\
        12  & $0.17 \pm 0.02$                   & $2.30 \pm 0.97$                    & $2.39 \pm 0.80$       & $51.95 \pm 3.95$            & $22.60 \pm 13.98$         & $816.44^\ast$             \\
        13  & $0.43 \pm 0.38$                   & $66.69 \pm 5.07$                   & $0.75 \pm 0.02$       & $54.55 \pm 0.17$            & $255.44 \pm 9.13$         & $0.09$               \\
        14  & $0.73 \pm 0.01$                   & $8.06 \pm 1.18$                    & $1.54 \pm 0.08$       & $52.04 \pm 0.43$            & $203.91 \pm 3.66$         & $1.35$               \\
        15  & $0.86 \pm 0.00$                   & $10.77 \pm 1.11$                   & $1.39 \pm 0.05$       & $52.39 \pm 0.25$            & $16.24 \pm 2.98$          & $0.38$               \\
        16  & $0.86 \pm 0.10$                   & $42.23 \pm 4.07$                   & $0.88 \pm 0.03$       & $54.14 \pm 0.09$            & $116.56 \pm 5.53$         & $0.15$               \\
        17  & $1.19 \pm 0.01$                   & $3.16 \pm 0.42$                    & $2.12 \pm 0.10$       & $43.55 \pm 1.53$            & $351.77 \pm 14.40$        & $13.72^\ast$              \\
        18  & $1.52 \pm 0.02$                   & $3.82 \pm 0.66$                    & $1.99 \pm 0.12$       & $42.88 \pm 2.14$            & $344.02 \pm 11.21$        & $15.60^\ast$              \\
        19  & $1.83 \pm 0.01$                   & $5.22 \pm 0.66$                    & $1.79 \pm 0.08$       & $44.33 \pm 1.34$            & $85.48 \pm 7.82$          & $3.92$               \\
        20  & $2.06 \pm 0.42$                   & $65.55 \pm 4.99$                   & $0.76 \pm 0.02$       & $53.83 \pm 0.20$            & $182.83 \pm 9.30$         & $0.10$               \\
        21  & $2.79 \pm 0.02$                   & $2.98 \pm 0.47$                    & $2.06 \pm 0.07$       & $30.22 \pm 2.86$            & $225.82 \pm 14.11$        & $16.62^\ast$              \\
        22  & $2.86 \pm 0.15$                   & $45.03 \pm 3.90$                   & $0.86 \pm 0.03$       & $52.88 \pm 0.19$            & $358.76 \pm 13.09$        & $0.28$               \\
        23  & $3.17 \pm 0.02$                   & $3.46 \pm 0.60$                    & $1.97 \pm 0.08$       & $30.51 \pm 3.18$            & $49.28 \pm 13.58$         & $17.30^\ast$              \\
        24  & $4.20 \pm 0.12$                   & $41.97 \pm 3.04$                   & $0.88 \pm 0.02$       & $51.81 \pm 0.23$            & $262.86 \pm 4.70$         & $0.08$               \\
        25  & $4.81 \pm 0.09$                   & $39.67 \pm 3.00$                   & $0.90 \pm 0.02$       & $51.18 \pm 0.28$            & $278.17 \pm 4.68$         & $0.09$               \\
        26  & $5.17 \pm 0.27$                   & $66.03 \pm 4.48$                   & $0.76 \pm 0.02$       & $52.45 \pm 0.20$            & $328.24 \pm 8.23$         & $0.07$               \\
        27  & $6.53 \pm 0.04$                   & $19.60 \pm 2.48$                   & $1.15 \pm 0.05$       & $44.86 \pm 1.28$            & $120.54 \pm 7.46$         & $0.99$               \\
        28  & $6.61 \pm 0.02$                   & $9.53 \pm 2.63$                    & $1.45 \pm 0.11$       & $34.65 \pm 5.09$            & $305.34 \pm 13.18$        & $19.93^\ast$              \\
        29  & $6.72 \pm 0.04$                   & $7.04 \pm 2.28$                    & $1.54 \pm 0.14$       & $29.16 \pm 6.46$            & $186.44 \pm 17.03$        & $47.63^\ast$              \\
        \hline
    \end{tabular}
\caption{\C nuclear spins localization results for NV 2.  Fit results for global parameters are $p_0=0.43\pm 0.02$, $\Tn=8.4\pm 0.76\unit{ms}$ and $\tread=1.86\pm0.24\unit{\us}$.  Errors for $\apar$, $\aperp$ and $\phi$ are calculated by bootstrapping. Errors for $r$, $\vartheta$, $\phi$ and the global parameters are calculated using Monte Carlo error propagation. The estimated uncertainty volume $\delta V$ (also depicted in Fig. 5f) is given in cubic Angstrom. $^\ast$ denote spins whose uncertainty volume is larger than the volume per carbon atom in the diamond lattice (5.69\,\AA$^3$).
    }
    \label{tbl:wm_loc_results_2}
\end{table}
\addtolength{\tabcolsep}{-2pt}
%


\clearpage
\section{Supplementary Figures}

\begin{figure}[h]
\includegraphics[width=0.4\textwidth]{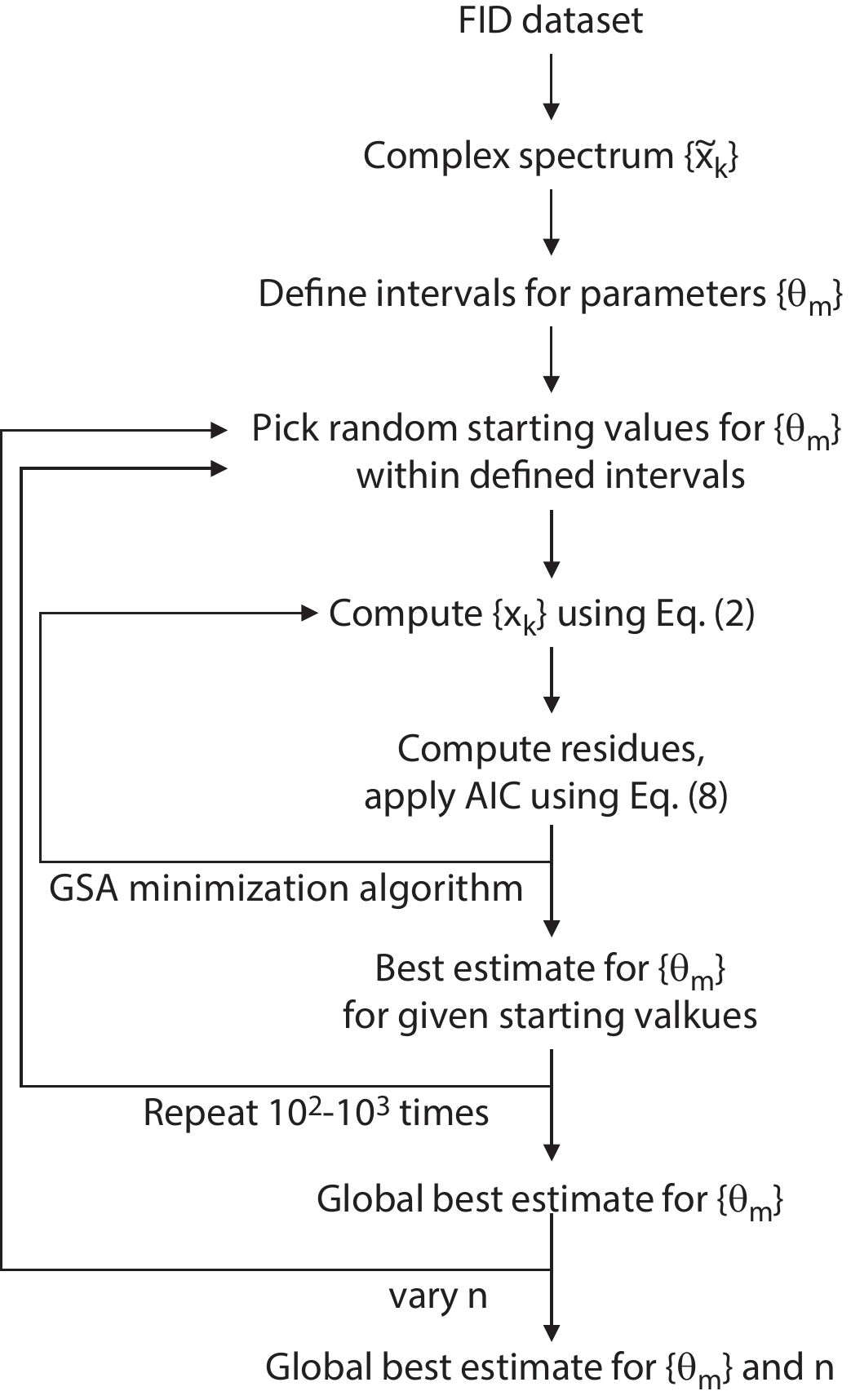}
    \caption[Flow diagram for maximum likelihood estimation.]{
    Flow diagram for maximum likelihood estimation, as discussed in Section VI of the main manuscript and Section~2. of the Supplemental Material.
    }
    \label{fig:flow_chart}
\end{figure}
%


%
\clearpage
\begin{figure}[h]
    \includegraphics[width=\textwidth]{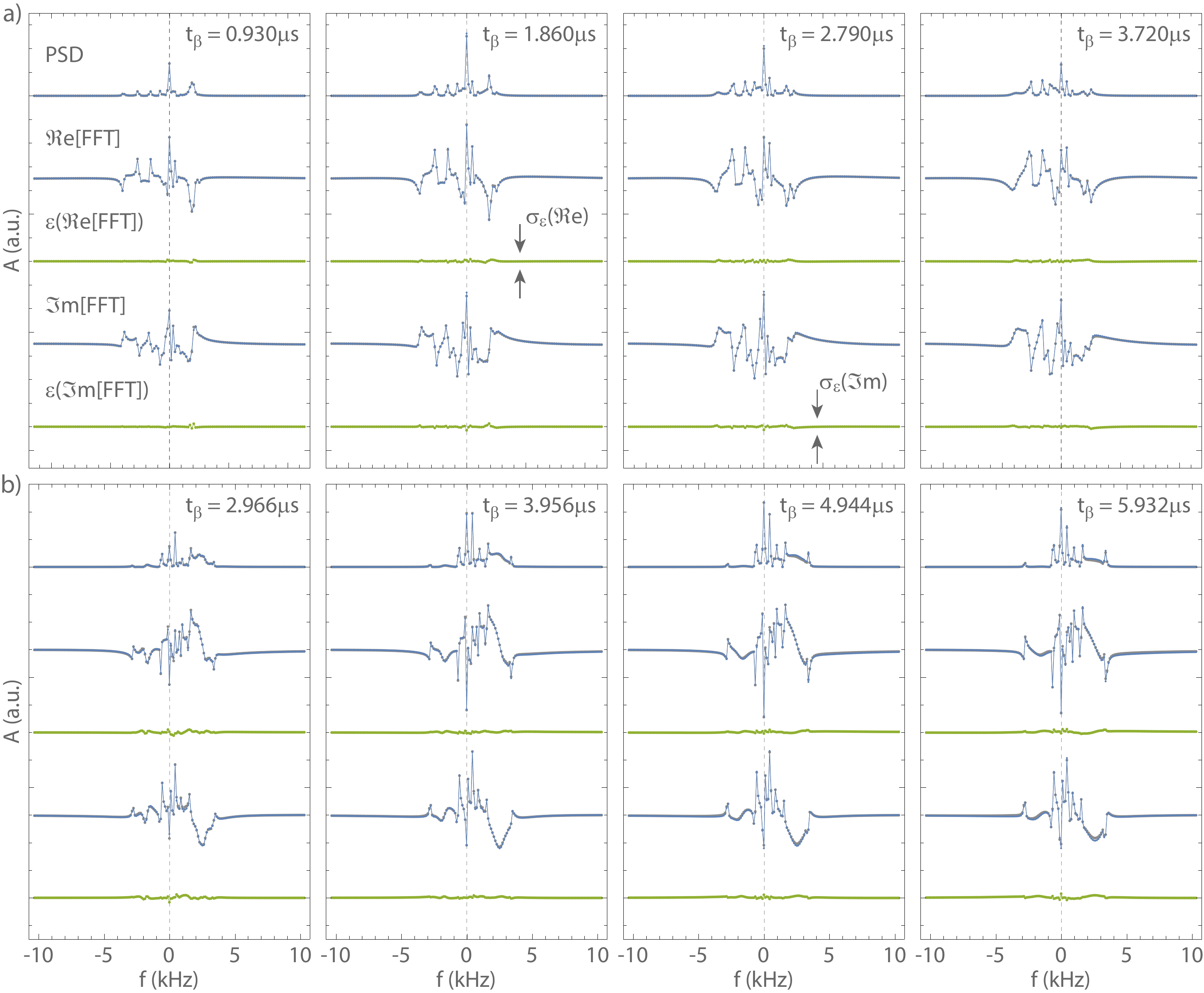}
    \caption[Second-order analytical model and numerical density-matrix simulations.]{
        \textbf{Second-order analytical model and numerical density-matrix simulations.}
        Complex Fourier spectra of the \C environment (\NVbns) for a series of interaction times $\tbeta$.  Shown are from top to bottom (vertically offset for clarity): power spectrum (PSD), real part of the complex spectrum ($\ReFFT$), fit residues for $\ReFFT$, imaginary part of the complex spectrum ($\ImFFT$), and fit residues for $\ImFFT$. The horizontal axis shows the spectral shift relative to the \C Larmor frequency. To illustrate the agreement between our analytical FID model (Eq.~2 in the main text) and the numerical density-matrix simulations, we take the measurement parameters (Table~\ref{tbl:params_fig2b}) and the best-fit parameters obtained for \NVb and simulate the resulting set of spectra (blue traces) using the density matrix formalism. The gray traces show the corresponding spectra obtained using our analytical model. The negligible residues (green traces), well smaller than those in Figs.4b,c ~in the main text, highlight the excellent match between the density-matrix simulations and our second-order analytical model.
    }\vfill
    \label{fig:density_matrix}
\end{figure}


%
\clearpage
\begin{figure}[h]
    \includegraphics[width=0.5\textwidth]{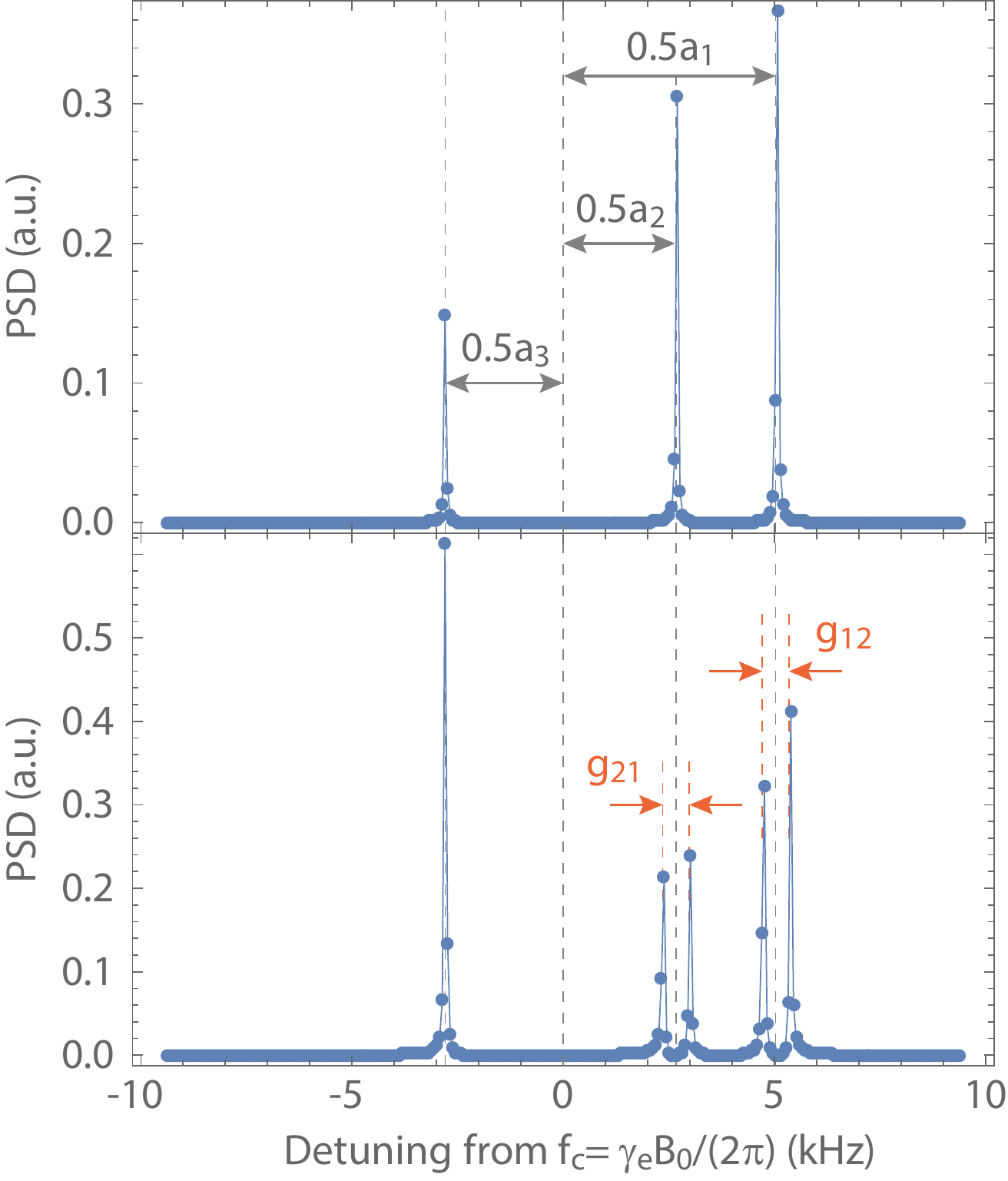}
    \caption[Simulation of weak measurement spectroscopy with non-vanishing inter-nuclear couplings.]{
        \textbf{Simulation of weak measurement spectroscopy with non-vanishing inter-nuclear couplings.}
        We consider three \C nuclear spins with the arbitrary hyperfine couplings $\apar=\cbracks{10.056,5.338,-5.593}\unit{kHz}$ and $\aperp=\cbracks{16.863,13.990,9.359}\unit{kHz}$ and nuclear \C-\C couplings $\cbracks{g_{12}=634.79,g_{13}=1.17,g_{23}=22.05}\unit{Hz}$.  The bias field is $B_0 = 202\unit{mT}$.  We simulate a time trace of $K=2,000$ weak measurements ($\tbeta=0.465\unit{\us}$) at a sampling time $\ts=8\unit{\us}$ and plot the power spectrum (blue dots) against the detuning from the detection frequency $\fc=1/(2\tau)$, where $\tau$ is the spacing between $\pi$ pulses (Fig.~2b in the main text).  In the absence of \C-\C couplings (top panel), we observe three peaks associated with the three nuclear spins.  The offset from $\fc$ is proportional to the parallel hyperfine coupling constant (grey arrows).  In contrast, for non-vanishing internuclear couplings (bottom panel), splittings proportional to $g_{ij}$ appear but only those within the spectral resolution $\delta f=1/(K\ts) = 156.25\unit{Hz}$ set by the Fourier limit of the time trace are resolved (orange arrows).
    }\vfill
    \label{fig:nuclear_couplings}
\end{figure}

\clearpage
\input{"references_supplementary.bbl"}

%% file: references.bbl
%

%% file: references_supplementary.bbl
%